\documentclass[aps,twocolumn,superscriptaddress]{revtex4-1}     
\usepackage{graphicx}
\usepackage{amsmath}
\usepackage{mathtools}
\usepackage{bm}
\usepackage{color}
\usepackage{hyperref}
\usepackage{dsfont}

\DeclareMathOperator{\Tr}{Tr}

\begin{document}
\renewcommand{\vec}[1]{\boldsymbol{#1}}
\newcommand{\up}{{\uparrow}}
\newcommand{\dw}{{\downarrow}}
\newcommand{\pa}{{\partial}}
\newcommand{\pd}{{\phantom{\dagger}}}
\newcommand{\bs}[1]{\boldsymbol{#1}}
\newcommand{\todo}[1]{{\textbf{\color{red}#1}}}
\newcommand{\eps}{{\varepsilon}}
\newcommand{\nn}{\nonumber}
\newcommand{\ie}{{\it i.e.},\ }
\def\eg{\emph{e.g.}\ }
\def\ea{\emph{et al.}}

\newcommand{\ls}[1]{ {\color{cyan} [#1]} }
\newcommand{\gj}[1]{ {\color{blue} [#1]} }
\newcommand{\sd}[1]{ {\color{magenta} [#1]} }

\title{Anisotropic exchange and non-collinear antiferromagnets on a noncentrosymmetric fcc structure as in the half-Heuslers}
\author{Seydou-Samba Diop}
\email{seydou-samba.diop@ens-lyon.fr}
\affiliation{Universit\'e de Lyon, \'{E}cole Normale Sup\'{e}rieure de
  Lyon, Universit\'e Claude Bernard Lyon I, CNRS, Laboratoire de physique, 46, all\'{e}e d'Italie, 69007 Lyon, France}
\author{George Jackeli}
\email{g.jackeli@fkf.mpg.de}
\altaffiliation{Also at Andronikashvili Institute of Physics, 0177 Tbilisi, Georgia}
\affiliation{Max Planck Institute for Solid State Research,
Heisenbergstrasse 1, D-70569 Stuttgart, Germany}
\affiliation{Institute for Functional Matter and Quantum Technologies, University of Stuttgart, Pfaffenwaldring 57, D-70569 Stuttgart, Germany}
\author{Lucile Savary}
\email{lucile.savary@ens-lyon.fr}
\affiliation{Universit\'e de Lyon, \'{E}cole Normale Sup\'{e}rieure de
  Lyon, Universit\'e Claude Bernard Lyon I, CNRS, Laboratoire de physique, 46, all\'{e}e
  d'Italie, 69007 Lyon, France}

\date{\today}

\begin{abstract}
One of the signatures of the face-centered cubic (fcc) antiferromagnet as a typical example of a geometrically frustrated system is the large ground state degeneracy of the classical nearest neighbor and next-nearest neighbor Heisenberg (isotropic) model on this lattice. In particular, collinear states are degenerate with non-collinear and non-coplanar ones: this degeneracy is accidental and is expected to be lifted by anisotropic exchange interactions. In this work, we derive the most general nearest and next-nearest neighbor exchange model allowed by the space-group symmetry of the noncentrosymmetric half-Heusler compounds, which includes three anisotropic terms: the so-called Kitaev, Gamma and Dzyaloshinskii-Moriya interactions -- most notably, the latter is allowed by the breaking of inversion symmetry in these materials and has not been previously been studied in the context of the fcc lattice. We compute the resulting phase diagram and show how the different terms lift the ground state degeneracy of the isotropic model, and lay emphasis on finding regimes where multi-q (non-collinear/non-coplanar) states are selected by anisotropy. We then discuss the role of quantum fluctuations and the coupling to a magnetic field in the ground state selection, and show that these effects can stabilize non-coplanar (triple-q) states. These results suggest that some half-Heusler antiferromagnets might host rare non-collinear/non-coplanar orders, which may in turn explain the unusual transport properties detected in these semimetals.
\end{abstract}

\maketitle

\section{Introduction}
\label{sec:intro}

True to the denomination of the face-centered cubic (fcc) lattice as a prototypical three-dimensional geometrically-frustrated system, a variety of magnetic orders and behaviors arise in model systems with this structure. For example, the phase diagram of the classical nearest- and next-nearest neighbor Heisenberg (isotropic) model on this lattice was established long ago and displays several antiferromagnetic ground states \cite{yamamoto_spin_1972, seehra_magnetic_1988, yildirim_frustration_1998, ader_magnetic_2001, gvozdikova_monte_2005, balla_degenerate_2020}. Among them, the commensurate states, dubbed type-I,II,III, all feature an accidental degeneracy between single-q and multi-q ground states \cite{yamamoto_spin_1972} (multi-q states are magnetic configurations with a superposition of symmetry-related ordering wavevectors). The addition of a magnetic field \cite{heinila_selection_1993, jackeli_frustrated_2004} and quantum fluctuations \cite{lefmann_quantum_2001, singh_spin_2017} lead to competition between these ground states and interesting excitation spectra. More recently, theoretical studies motivated by double perovskites showed that the incorporation of nearest-neighbor anisotropic terms (including the bond-dependent Kitaev interaction \cite{jackeli_mott_2009, chaloupka_kitaev-heisenberg_2010, kimchi_kitaev-heisenberg_2014}) in the classical model leads to yet new magnetic phases, including incommensurate spiral states \cite{ishizuka_magnetism_2014, cook_spin-orbit_2015, li_kitaev_2017, revelli_spin-orbit_2019}.

\begin{figure}[h]
\begin{center}
\includegraphics[width=8.5cm]{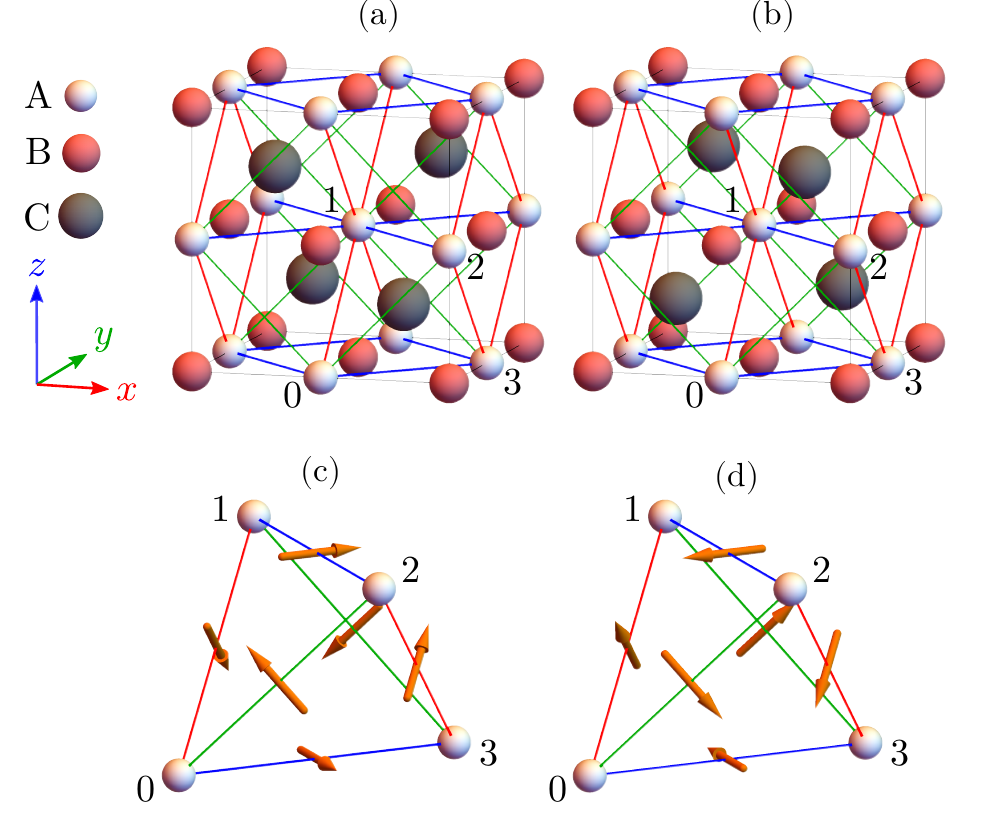}
\caption{(a, b) Two enantiomer crystal structures of the half-Heusler compound ABC. They differ by the location of the C-sites which break the inversion-symmetry. The nearest-neighbor bonds are shown in green, blue and red according to their $\gamma$-axes $x$, $y$ and $z$ respectively. (c, d) Dzyaloshinskii-Moriya vectors $\mathbf{D}_{ij}$ for the bonds of an elementary tetrahedron. The four sites are labeled by $i,j \in \{0,1,2,3\}$ and the bonds are oriented as $i \rightarrow j$ with the convention $i<j$. The two inequivalent configurations of DM vectors (c, d) are related by $\mathbf{D}_{ij} \rightarrow - \mathbf{D}_{ij}$ and can be associated with the two enantiomers (a, b). }
\label{fig:kitaev_dm}
\end{center}
\end{figure}

Experimentally, many fcc-based compounds, such as MnS$_2$/MnTe$_2$ \cite{hastings_antiferromagnetic_1959}, MnO \cite{bloch_order-parameter_1974}, double-perovskites \cite{aczel_frustration_2013}, and Heuslers/half-Heuslers \cite{canfield_magnetism_1991} were discovered to exhibit antiferromagnetic order. In this work we will be more particularly interested in the half-Heusler antiferromagnets, with chemical formula ABC and a crystal structure composed of three interpenetrating fcc sublattices, see Fig.~\ref{fig:kitaev_dm}. Typically, one of the fcc sublattices is occupied by rare-earth ions which carry magnetic moments of localized $4f$ electrons. In turn, the half-Heusler family provides many candidate materials to study anisotropic exchange and the resulting novel behaviors. Moreover, unlike the other aforementioned materials, the magnetic sublattice of rare-earth ions is embedded in an environment with lower symmetry than that of the isolated fcc lattice: most notably, it breaks inversion symmetry. As we show in this manuscript, this allows for the existence of Dzyaloshinskii-Moriya interactions which have, to the best of our knowledge, not been previously discussed in the literature for the fcc lattice.

Neutron diffraction has shown that type-I and type-II antiferromagnetic orders appear experimentally in some half-Heuslers of the subfamilies RPtBi \cite{wosnitza_magnetic-field-_2006, muller_magnetic_2014, muller_magnetic_2015, suzuki_large_2016, singha_magnetotransport_2019, sukhanov_magnon_2020, zhang_field-induced_2020} and RPdBi \cite{nakajima_topological_2015, pavlosiuk_antiferromagnetism_2016, pavlosiuk_magnetic_2018}, where R denotes the rare-earth (lanthanide) element. The trend appears to be the following: type-I order for lighter R (Nd, Ce) and type-II order for heavier R (Sm to Lu), see Table~\ref{tab:experimental}. However such experiments cannot unequivocally differentiate single-q and multi-q arrangements of a given ordering type (I or II). For example it is unclear how to identify, with the Bragg spectrum alone, the difference between large-scale multi-q order, and the existence of multiple domains of single-q states, i.e.\ macroscopic configurations where the symmetry-related ordering wavevectors appear in multiple real-space domains. Moreover, since these measurements are typically made in the presence of an external magnetic field, the observed orders need not a priori correspond to the zero-field ground states. This motivates an extensive theoretical investigation of the allowed magnetic states in the half-Heuslers, and in particular whether multi-q states can be realized and stabilized in these compounds.

\begin{table}[tb]
\centering
\begin{tabular}{lccc}
\hline\hline
  Compound & Magnetic order & $T_N$(K) & Reference \\
\hline
CePtBi & Type-I & 1.15 & \cite{wosnitza_magnetic-field-_2006} \\
NdPtBi & Type-I & 2.18 &  \cite{muller_magnetic_2015} \\
\hline
GdPtBi & Type-II & 9.2 & \cite{muller_magnetic_2014, suzuki_large_2016, sukhanov_magnon_2020} \\
GdPdBi & Type-II & 12.8 &  \cite{nakajima_topological_2015, pavlosiuk_magnetic_2018} \\
SmPdBi & Type-II & 3.2 &  \cite{nakajima_topological_2015} \\
TbPtBi & Type-II & 3.4 & \cite{singha_magnetotransport_2019} \\
TbPdBi & Type-II & 5.1 &  \cite{nakajima_topological_2015, pavlosiuk_magnetic_2018} \\
DyPtBi & Type-II & 3.5 & \cite{zhang_field-induced_2020} \\
DyPdBi & Type-II & 3.5 &  \cite{nakajima_topological_2015, pavlosiuk_magnetic_2018} \\
HoPdBi & Type-II & 1.9 &  \cite{nakajima_topological_2015, pavlosiuk_antiferromagnetism_2016, pavlosiuk_magnetic_2018} \\
ErPdBi & Type-II & 1.1  &  \cite{nakajima_topological_2015, pavlosiuk_magnetic_2018} \\
\hline\hline
\end{tabular}
\caption{Zero-field magnetic orders and their N\'{e}el temperature, $T_N$, in half-Heusler antiferromagnets reported in the experimental literature.}
\label{tab:experimental}
\end{table}

The purpose of this study is therefore multifold: {\em (i)} derive the spin bilinear symmetry-allowed exchange Hamiltonian up to second neighbor interactions including anisotropic terms (and including a Dzyaloshinskii-Moriya interaction), {\em (ii)} determine the classical phase diagram of the resulting multi-parameter model, which includes many spiral states, {\em (iii)} identify the dominant mechanisms which stabilize multi-q states over the single-q ones, namely anisotropy, magnetic field, or quantum fluctuations. We indeed lay particular emphasis on the search for parameter regimes where noncollinear and noncoplanar antiferromagnetic orders, such as \emph{multi-q} states, are stable. Multi-q states are notoriously rare in classically frustrated systems. Indeed, not only are they often forbidden by the fixed-norm constraint of the spin, $|\mathbf{S}|={\rm const}$, but even when they are allowed, fluctuations (thermal and quantum) in {\em isotropic} models are expected to lift the degeneracy in favor of (collinear) single-q states \textit{via} the so-called ``order-by-disorder'' mechanism \cite{villain_order_1980, henley_ordering_1987, henley_ordering_1989, sheng_ordering_1992, schick_quantum_2020}. Here, however, as mentioned earlier, the fcc lattice the classical Heisenberg model does allow for an accidental degeneracy between single-q and multi-q ground states \cite{yamamoto_spin_1972}, and our model is highly anisotropic.

Moreover, beyond the scarcity of such orders in the context of magnetism, the role of noncollinear antiferromagnetism in inducing an anomalous Hall effect (AHE) may be important in the half-Heusler compounds, most of which are itinerant systems.  Indeed, it has been suggested that the intrinsic AHE observed in some spin-orbit coupled non-collinear antiferromagnets, such as the tetragonal compounds Mn$_3$Sn \cite{nakatsuji_large_2015} and Mn$_3$Ge \cite{nayak_large_2016}, can be interpreted as resulting from real-space Berry phases acquired by itinerant electrons coupled to a background of noncollinear magnetic moments \cite{chen_anomalous_2014, zhang_real-space_2020}. The recent discovery of a large instrinsic AHE in several half-Heusler antiferromagnets such as GdPtBi \cite{suzuki_large_2016, hirschberger_chiral_2016, shekhar_anomalous_2018}, TbPtBi \cite{singha_magnetotransport_2019}, DyPtBi \cite{zhang_field-induced_2020} and DyPdBi \cite{mukhopadhyay_electronic_2019} thus raises the question of the favored magnetic orders in this class of materials and begs for further understanding of the link between non-collinear magnetism and anomalous transport. While the half-Heuslers have mostly attracted attention because of other low-temperature properties such as (unconventional) superconductivity \cite{pan_superconductivity_2013, nakajima_topological_2015} and topological phases \cite{chadov_tunable_2010, yan_half-heusler_2014, hirschberger_chiral_2016}, noncollinear magnetism, and a noncollinear-induced anomalous Hall effect would provide yet further exciting physics in this large family of compounds.

The remainder of this manuscript goes as follows. We first derive the most general quadratic exchange Hamiltonian allowed by the symmetries of the half-Heusler crystal structure.  We find that, in addition to the nearest-neighbor Heisenberg interaction $J_1$, and the two nearest-neighbor anisotropic terms discussed previously in the literature, namely the bond-dependent Ising-type $K$ and the symmetric off-diagonal $\Gamma$ interactions (in the literature often referred to as Kitaev and `Gamma' couplings, respectively), a Dzyaloshinskii-Moriya interaction $D$ is also allowed in a noncentrosymmetric environment, as is present in the half-Heuslers (Sec.~\ref{sec:model}). At the next-nearest neighbor level, only Heisenberg $J_2$ and Kitaev interactions are allowed.

We then establish the classical ground state phase diagram of the anisotropic model. To this end, we first compute phase diagrams \textit{via} the Luttinger-Tisza method (Sec.~\ref{sec:LT_diagrams}), which gives access to the stability regions of the different ordering wavevectors in the $J_1$-$J_2$-$K$-$\Gamma$-$D$ parameter space. There, we find that the type-I, type-II and type-III orders identified in the isotropic model extend to large regions of the phase diagram when anisotropy is included, and that a flurry of incommensurate phases appear as well. Since the Luttinger-Tisza approach only considers the single-wavevector configurations, this method alone cannot establish whether the superposition of symmetry-related wavevectors in the form of a multi-q state is energetically favorable or not compared with the single-q configuration, and we thus resort to a complementary analysis to unambiguously determine the ground states. Therefore in Sec.~\ref{sec:energetics}, we investigate whether anisotropic interactions lift the degeneracies between single-q and multi-q states in the type-I, II, III phases by explicitly comparing the anisotropy energy in these states.
In particular we show that, although the Kitaev and Gamma couplings alone preserve some of the accidental degeneracies of the $J_1$-$J_2$ Heisenberg model, the combination of anisotropy and an external magnetic field select non-collinear and non-coplanar states (Sec.~\ref{sec:field}) for some directions of the magnetic field. The Dzyaloshinskii-Moriya interaction itself drives the system to a single-q, but {\em non-collinear} state.

In regions of parameter space where an accidental degeneracy remains at the classical level, the ground state is likely to be selected by the effect of fluctuations. In Sec.~\ref{sec:fluctuations}, we show how quantum fluctuations lift these degeneracies, selecting single-q states in the limit of small anisotropy, but driving the system towards multi-q states in some stronger anisotropy regimes. In some of the noncollinear states, one may expect an induced anomalous Hall effect in systems which contain itinerant electrons as well.



\section{Exchange Hamiltonian}
\label{sec:model}

\subsection{Derivation from crystal symmetries}

We first derive and consider the most general form of nearest and next-nearest neighbor exchange interactions on an fcc sublattice embedded in the {\em noncentrosymmetric} space group F$\bar{4}$3m (no.~216, with tetrahedral point group $T_d$), associated with the crystal structure of the half-Heusler compounds depicted in Fig.~\ref{fig:kitaev_dm}(a,b) \cite{yan_half-heusler_2014} whose generic chemical formula is ABC. The A-sites (typically rare-earth ions which carry a magnetic moment) and the B-sites occupy two fcc sublattices and form a rocksalt-type structure while the C-sites sit on an fcc lattice which breaks the inversion symmetry of the structure and give rise to a Dzyaloshinskii-Moriya interaction as we discuss below. Two locations are possible for the C-sites: the centers of the up-tetrahedra or those of the down-tetrahedra formed by four nearest-neighbor A-sites. This gives two twin crystals, shown in Fig.~\ref{fig:kitaev_dm}(a) and Fig.~\ref{fig:kitaev_dm}(b).

We consider a bilinear exchange Hamiltonian
\begin{equation}
\label{eq:H1H2}
  H = H_1 + H_2
\end{equation}
for the magnetic moments on the A-sites, which includes interactions between nearest ($H_1$) and next-nearest ($H_2$) neighbors, and find that the most general symmetry-allowed forms are
\begin{equation} \label{eq:H1}
\begin{split}
H_1 &= J_{1} \sum_{\langle i,j \rangle } \mathbf{S}_{i} \cdot \mathbf{S}_{j} + K \sum_{\langle i,j \rangle _{\gamma}} S^{\gamma}_{i} S^{\gamma}_{j} \\
&+ \Gamma \sum_{\langle i,j \rangle _{\gamma}} \xi_{ij} (S^{\alpha}_{i} S^{\beta}_{j}  + S^{\beta}_{i} S^{\alpha}_{j}  )\\
&+ \sum_{\langle i,j \rangle }\textbf{D}_{ij} \cdot (\mathbf{S}_{i} \times \mathbf{S}_{j}),
\end{split}
\end{equation}
and
\begin{equation}
  \label{eq:H2_1}
  \begin{split}
    H_2 &= J_{2} \sum_{\langle\!\langle i,j \rangle\!\rangle } \mathbf{S}_{i} \cdot \mathbf{S}_{j} + K_{2} \sum_{\langle\!\langle i,j \rangle\!\rangle _{\gamma}} S^{\gamma}_{i} S^{\gamma}_{j} .
  \end{split}
\end{equation}
In the rest of the manuscript we consider only the isotropic Heisenberg interaction for second neighbors, i.e.\ we take
\begin{equation}
\label{eq:H2_2}
H_2 \rightarrow  J_{2} \sum_{\langle\!\langle i,j \rangle\!\rangle} \mathbf{S}_{i} \cdot \mathbf{S}_{j}.
\end{equation}

For nearest neighbors, besides the Heisenberg interaction with strength $J_1$, three anisotropic couplings are allowed:

{\em (i)} the Kitaev interaction (with strength $K$) couples the components of the spins along the bond-dependent Ising-like axes $\gamma\in\{x,y,z\}$, where the $\gamma$ index labels the cubic plane in which the bonds lie. Note that there are six inequivalent nearest-neighbor bonds, which come in pairs of $90^{\circ}$-rotated ones, both carrying a $\gamma$ index. Both types of inequivalent $\gamma$ bonds lie in the plane perpendicular to the $\gamma$-axis, as depicted in Fig.~\ref{fig:kitaev_dm}. Each 0123 tetrahedron contains exactly one copy of each inequivalent bond, as shown Fig.~\ref{fig:kitaev_dm}. In the case of {\em next}-nearest neighbors, the Kitaev axes correspond to the axes $x$, $y$, $z$ along which the bonds are oriented. 

{\em (ii)} A symmetric off-diagonal term, with strength $\Gamma$ which involves the bond-dependent axes $\alpha$ and $\beta$ which are orthogonal to the Kitaev axis $\gamma$ of the bond. Inequivalent bonds of the same $\gamma$ type, carry a sign $\xi_{ij} = +1$ for the bonds $\langle 0, 1 \rangle$, $\langle 0, 2 \rangle$ and $\langle 0, 3 \rangle$, and $\xi_{ij} = -1$ for their $90^{\circ}$-rotated counterparts $\langle 2,3 \rangle$, $\langle 1,3 \rangle$ and $\langle 1,2 \rangle$. As an example, for the bond $\langle 0,3 \rangle$ which lies in the $xy$ plane, the symmetric part of the exchange interaction is of the form $J_1 \mathbf{S}_0 \cdot \mathbf{S}_3 + K S_0^{z}S_3^{z} + \Gamma (S_{0}^{x} S_{3}^{y} + S_{0}^{y} S_{3}^{x})$.

 {\em (iii)} An antisymmetric, i.e.\ Dzyaloshinskii-Moriya (DM) coupling, which is allowed because, in the half-Heusler structure, the bond centers are not inversion centers, in contrast to the case of the isolated fcc lattice. The DM vector $\textbf{D}_{ji} = -\textbf{D}_{ij}$ with norm $D$ associated with a bond $\langle i,j\rangle$ is orthogonal to the bond direction $\hat{\mathbf{r}}_{ij}=(\mathbf{r}_j - \mathbf{r}_i)/|\mathbf{r}_j - \mathbf{r}_i|$ and to the unit vector $\hat{\mathbf{e}}_{\gamma} \in \{\hat{\mathbf{e}}_{x}, \hat{\mathbf{e}}_{y}, \hat{\mathbf{e}}_{z}\}$ corresponding to the Kitaev axis of the bond, i.e.:
\begin{equation}
\label{eq:dm_vectors}
\mathbf{D}_{ij} = D \xi_{ij} ~ \hat{\mathbf{r}}_{ij} \times \hat{\mathbf{e}}_{\gamma}.
\end{equation}
This configuration of DM vectors is plotted in Fig.~\ref{fig:kitaev_dm}(c). 
Note that, in Eq.~\eqref{eq:dm_vectors}, a global sign inversion $D\rightarrow -D$ gives another configuration of DM vectors which respects the lattice symmetries, as shown in Fig.~\ref{fig:kitaev_dm}(d). These two allowed configurations can be associated with the two choices for the location of the C-sites which break the inversion symmetry of the structure Fig.~\ref{fig:kitaev_dm}(a,b). Note that the DM configurations are similar on the tetrahedra of the pyrochlore lattice \cite{elhajal_ordering_2005, canals_ising-like_2008}.

It is noteworthy that the Kitaev and Gamma anisotropies are directly allowed by the symmetries of the pure fcc lattice, and thus may be expected in most fcc magnets with strong spin-orbit coupling, such as rare-earth half- and full-Heusler compounds, and double perovskites \cite{cook_spin-orbit_2015, romhanyi_spin-orbit_2017, li_kitaev_2017, revelli_spin-orbit_2019}. In contrast, the DM term is only allowed in half-Heusler or zinc blende-like structures where inversion symmetry is broken by the C sublattice. The Gamma and DM anisotropies are forbidden for next-nearest neighbors, because of the $C_2$-rotation symmetry around the three axes $x$, $y$, $z$. One can expect that the heavier the rare-earth element, the larger the spin-orbit coupling and thus the larger the anisotropic exchange terms. Given the numerous possible of element substitutions, we expect that many values of the $K/J_1$, $\Gamma/J_1$, $D/J_1$ ratios can be realized in the half-Heusler family.

In the next subsection, we briefly summarize the known results for the isotropic $J_1$-$J_2$ Heisenberg model, with Appendix~\ref{sec:heisenberg}
providing a more detailed review. Afterwards, we discuss the effects of the anisotropic terms $K$, $\Gamma$, and $D$ on the phase diagram of the multiparameter model, Eqs.~(\ref{eq:H1H2}-\ref{eq:H2_2}).

\subsection{Review of the $J_1$-$J_2$ isotropic model}
\label{sec:review-j_1-j_2}

The phase diagram of the isotropic $J_1$-$J_2$ model contains
three commensurate antiferromagnetic orders for $J_1 > 0$: 
type-I order (with ordering wavevector $\mathbf{q}=(\pi,0,0)$, antiferromagnetic modulation between neighboring [100] planes, stable when $J_2/J_1 < 0$), type-II order ($\mathbf{q} = (\pi,\pi,\pi)/2$, antiferromagnetic modulation between neighboring [111] planes, favored when $0< J_2/J_1 < 1/2$) and type-III order ($\mathbf{q} = (\pi,\pi/2,0)$, stable when $J_2/J_1 > 1/2$). All $\mathbf{q}$ vectors are expressed in units of $2/a$, where $a$ is the cubic unit cell parameter.
For $J_2=0$, all spiral states with wavevectors of the form $\mathbf{q}=(\pi,Q,0)$, with $Q\in\mathds{R}$, are ground states: in particular, type-I order ($Q=0$) is degenerate with type-III order ($Q=\pi/2$). In the literature, these orders are sometimes labeled by the position of the corresponding wavevector in the fcc Brillouin zone, namely $X$, $L$, and $W$ for type-I, type-II and type-III, respectively.

For each of these orders, the spins can be parametrized in the following form:
\begin{equation}
\label{eq:general_S_i}
\mathbf{S}_i = \frac{1}{2} \sum_{\ell}\bigl[ \mathbf{u}_{\ell} e^{i \mathbf{q}_{\ell} \cdot \mathbf{r}_i}+\mathbf{u}^{*}_{\ell} e^{-i \mathbf{q}_{\ell} \cdot \mathbf{r}_i}\bigr],
\end{equation}
where the $\mathbf{q}_{\ell}$ are the ordering wavevectors related by cubic symmetry -- for example $\mathbf{q}_1 = (\pi,0,0)$, $\mathbf{q}_2 = (0,\pi,0)$, and $\mathbf{q}_3 = (0,0,\pi)$ for type-I order -- and the $\mathbf{u}_{\ell}$ are vectors chosen such that $|\mathbf{S}_i| = 1$ \cite{yamamoto_spin_1972}. We call a single-q state a configuration where only one of the $\mathbf{u}_{\ell}$ vectors is non-zero, a double-q state a configuration with two non-zero vectors, etc. It is sometimes useful to also consider a subset of the manifold Eq.~(\ref{eq:general_S_i}) made of \emph{equal-weight} states, where the non-zero $\mathbf{u}_{\ell}$ vectors have equal magnitude (that is, $|\mathbf{u}_\ell|=1,1/\sqrt{2},1/\sqrt{3}$, respectively, for the single-q, double-q and triple-q states). As mentioned, the ground state manifold of the $J_1$-$J_2$ model gives an accidental degeneracy between the collinear single-q states, and multi-q states \cite{yamamoto_spin_1972} throughout the phase diagram, which can be built to be noncollinear and even noncoplanar. Namely,
\begin{itemize}
\item type I: collinear single-q, non-collinear double-q, non-coplanar triple-q ;
\item type II: collinear single-q, non-collinear double-q, non-coplanar triple-q, collinear/non-collinear/non-coplanar quadruple-q;
\item type III: collinear/non-collinear single-q, non-collinear/non-coplanar double-q, non-coplanar triple-q.
\end{itemize}
In Appendix~\ref{sec:heisenberg}, we give a detailed description of the three commensurate AFM phases and plot the 
corresponding spin arrangements. Note that an extensive study of the fcc Heisenberg (isotropic) model up to third neighbor exchange $J_3$, including the parametrization of the spin vectors in the AFM states can be found in Ref.~\cite{balla_degenerate_2020}.
In what follows, we explore how the anisotropic $K$, $\Gamma$, and $D$ couplings, Eq.~\eqref{eq:H1}, 
fully or partly lift the degeneracy between single-q and multi-q states.

\section{Luttinger-Tisza phase diagrams}
\label{sec:LT_diagrams}

To determine the favored orders in the multiparameter classical model Eq.~\eqref{eq:H1H2}, we first use the Luttinger-Tisza approach. This method determines the values of $\mathbf{q}$ that minimize the Fourier transform of the exchange interaction \cite{litvin_luttinger-tisza_1974}. Physically, this provides the ground states of the system if they can be described by a single ordering wavevector $\mathbf{q}$. In Section \ref{sec:energetics} we will explicitly compare in our full model the energies of the thereby determined single-q states, with those of the multi-q states that are degenerate {\em in the isotropic case}.

\begin{figure*}[htbp]
\begin{center}
\includegraphics[width=1 \textwidth]{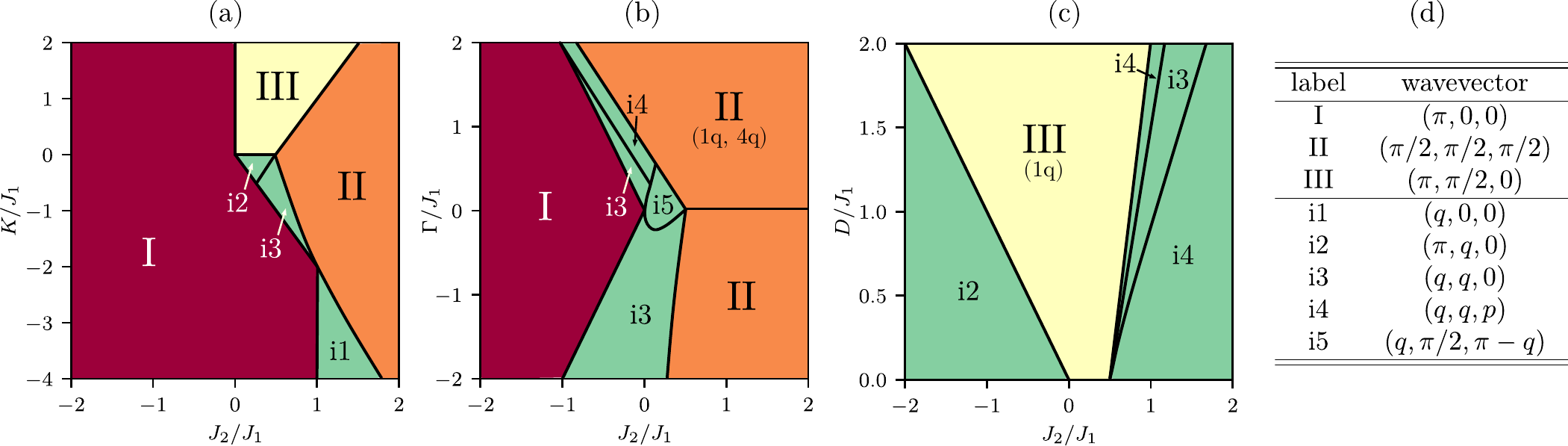}
\caption{Luttinger-Tisza phase diagrams for the anisotropic model with $J_1>0$. We consider the $j_1-J_2$ Heisenberg interactions and we vary separately the relative strength of: (a) the Kitaev coupling, at $\Gamma,D=0$, (b) the Gamma coupling, at $K,D=0$, (c) the Dzyaloshinskii-Moriya coupling, at $K,\Gamma=0$. In some regions of the diagrams, anisotropy lifts the degeneracy between single-q and multi-q states (as described in Sec.~\ref{sec:energetics}). In this case we indicate the corresponding ground state configuration (`1q', `2q', ...). (d) Luttinger-Tisza wavevectors associated with the different phases: the commensurate AFM orders I, II, III, and the incommensurate spiral phases (labeled `i1', ..., `i5'). Here $q$ and $p$ are fixed and unique (up to lattice symmetries) but vary continuously within the incommensurate phases with the parameters of the model. All wavevectors are expressed in units of $2/a$, where $a$ is the cubic unit cell parameter.}
\label{fig:LT_diagrams}
\end{center}
\end{figure*}

\begin{figure*}[htbp]
\begin{center}
\includegraphics[width=1 \textwidth]{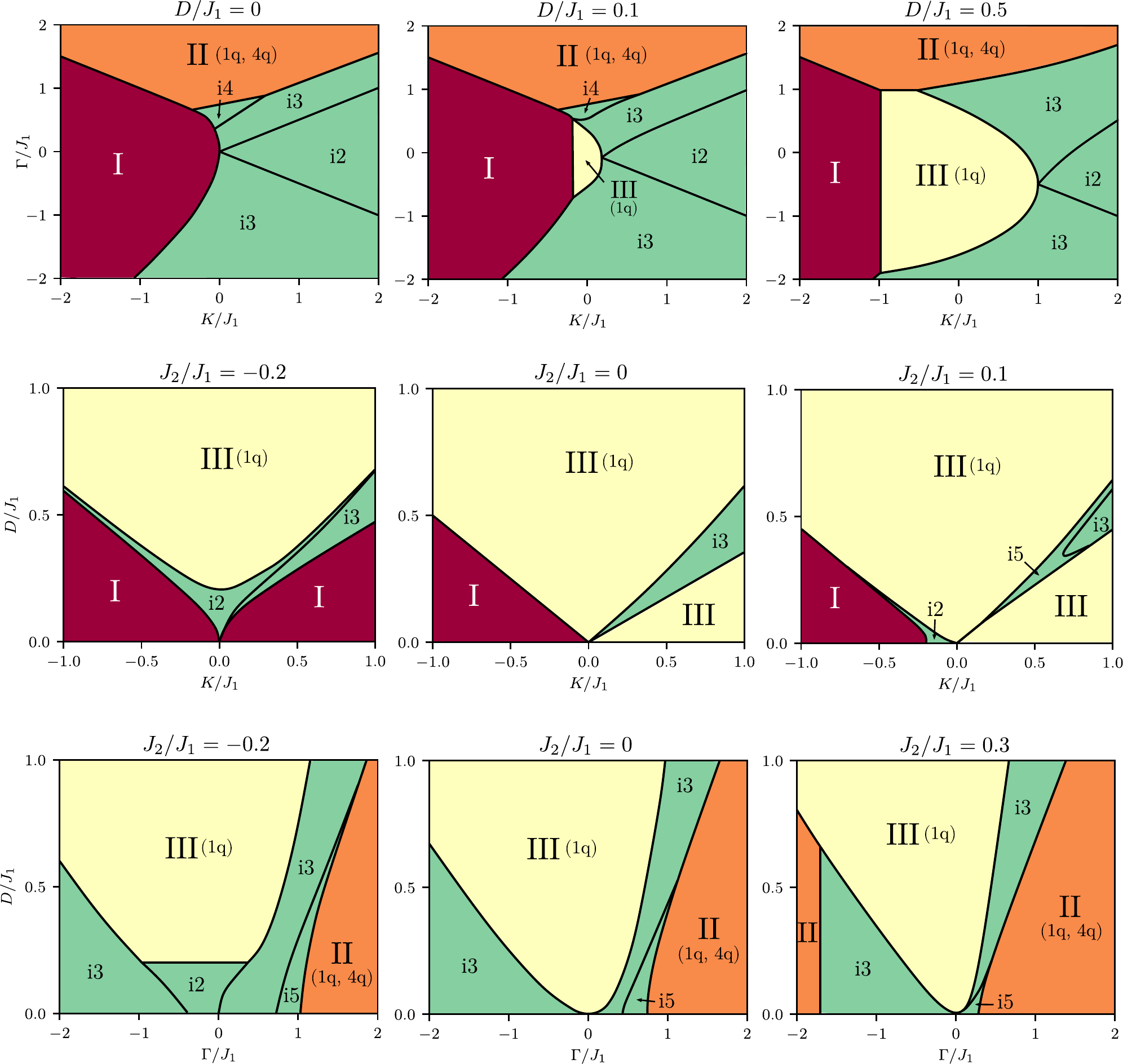}
\caption{Selection of cuts of the Luttinger-Tisza phase diagram in the five-dimensional parameter space: $(K,\Gamma)$ plane (upper row), $(K,D)$ plane (middle row) and in the $(\Gamma, D)$ plane (bottom row). For the $(K,\Gamma)$ plane, several values of the DM coupling $D$ were considered. For the $(K, D)$ and $(\Gamma, D)$ planes, three different values of $J_2$ were chosen, namely ferromagnetic (left), vanishing (middle), antiferromagnetic (right). The phases are labeled like in Fig.~\ref{fig:LT_diagrams}.}
\label{fig:LT_diagrams_2}
\end{center}
\end{figure*}

In the Luttinger-Tisza approach, we start by writing the Hamiltonian in Fourier space as 
\begin{equation}
\begin{split}
H &= \frac{1}{2} \sum_{i,j} \mathbf{S}_i A_{ij} \mathbf{S}_j = \frac{1}{2} \sum_{\mathbf{q}} \mathbf{S}_{-\mathbf{q}} A(\mathbf{q}) \mathbf{S}_{\mathbf{q}}, \\
\mathbf{S}_{\mathbf{q}} &= \frac{1}{\sqrt{N}} \sum_i \mathbf{S}_i e^{-i\mathbf{q} \cdot \mathbf{r}_i} \\ 
A(\mathbf{q}) &= \sum_{j} A_{ij} e^{i\mathbf{q}\cdot (\mathbf{r}_j - \mathbf{r}_i)},
\end{split}
\end{equation}
and we define the Luttinger-Tisza wavevectors as those which minimize the lowest eigenvalue of the Hermitian matrix $A(\mathbf{q})$. In Fig.~\ref{fig:LT_diagrams} we show three cuts in the phase diagram of $H$, assuming $J_1 >0$, obtained by varying the ratio $J_2/J_1$ and one of the three anisotropic couplings, while setting the other two to zero. The diagrams show that the the type-I, II and III AFM phases discussed above are the only stable {\em commensurate} orders and that a variety of incommensurate spiral states also exist (the green regions in Fig.~\ref{fig:LT_diagrams}). In these incommensurate phases, the wavelength of the modulation varies continuously with the coupling parameters and does not coincide with an integer number of lattice spacings, except for fine-tuned parameters. More precisely, the wavevectors describing the incommensurate spiral phases vary along the high-symmetry lines of the fcc Brillouin zone. In the rest of this work, we will focus on the commensurate AFM orders, I, II and III.

In Fig.~\ref{fig:LT_diagrams_2} we plot the Luttinger-Tisza phase diagrams of the model along several cuts in the five-dimensional parameter space. In particular, in the first row of Fig.~\ref{fig:LT_diagrams_2} we investigate how the $(K,\Gamma)$ phase diagram with $J_2$ (obtained in earlier works \cite{cook_spin-orbit_2015, revelli_spin-orbit_2019} with the Luttinger-Tisza method and confirmed by Monte-Carlo simulations) is modified when we add the Dzyaloshinskii-Moriya coupling $D$. We see that a region appears near the center of this diagram, where the ground state has type-III order. Similarly, in the bottom two rows we plot the cuts in the $(K,D)$ and $(\Gamma, D)$ planes, for several values of $J_2$. These diagrams display the richness of the multi-parameter model, which hosts many transitions between commensurate (type-I, II, and III) and incommensurate spiral orders.

\section{Single-q versus multi-q states: anisotropy energy comparison}
\label{sec:energetics}

\begin{figure}[htbp]
\begin{center}
\includegraphics[width=\columnwidth]{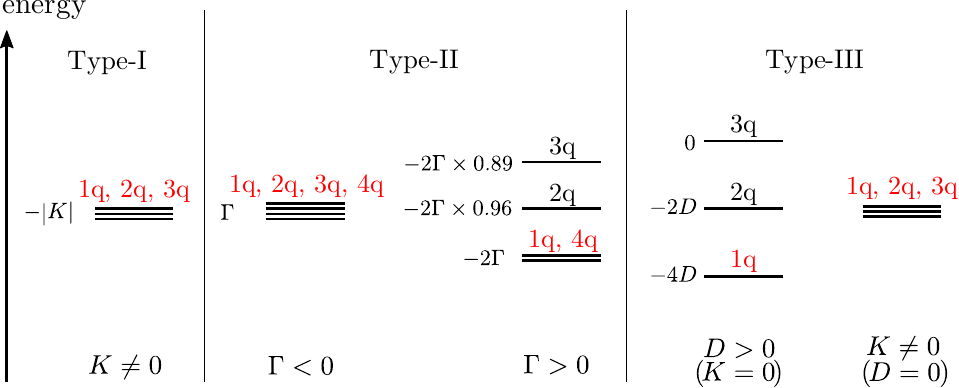}
\caption{Schematic representation of the ground states and degeneracies of the model, in the presence of anisotropy. In each case, the `1q', `2q', `3q' and `4q' levels represents only the lowest-energy state of the \emph{equal-weight} single-q, double-q, triple-q and quadruple-q manifold, respectively.
}
\label{fig:energy levels}
\end{center}
\end{figure}

The Luttinger-Tisza analysis presented in the previous section gives us an overview of the phases stabilized by the different parameters of the classical model, and the transition points between these phases. However, because this method only gives access to the ordering wavevector without enforcing the length constraints $|\mathbf{S}_i| = 1$ explicitly, it does not allow us to unambiguously determine the favored spin arrangement in each region of the diagram when there exist multiple degenerate wavevectors, i.e.\ in particular whether the favored state is single-q or multi-q. We therefore now supplement the Luttinger-Tisza analysis, by minimizing explicitly the anisotropy energy with the length constraints. In this section we compute the expectation value of the anisotropic interaction terms in the three phases I, II and III. This method allows us to compare the energy of our single-q, double-q and triple-q variational Ans\"atze, which are all ground states of the isotropic model. In Fig.~\ref{fig:energy levels} we summarize the results of the following subsections.

\subsection{Degeneracy in the type-I phase}
\label{sec:degeneracy-type-i}

In this subsection, we consider parameters of the model where the type-I states are energetically favored in the Luttinger-Tisza approach. Type-I order is defined by the symmetry-related ordering wavevectors $\mathbf{q}_1=(\pi,0,0)$, $\mathbf{q}_2=(0,\pi,0)$ and $\mathbf{q}_3=(0,0,\pi)$, so that in these phases the spins take the form:
\begin{equation} \label{eq:spins_I}
\textbf{S}_{i} = \mathbf{u}_1 e^{i \mathbf{q}_1 \cdot \textbf{r}_i} + \mathbf{u}_2 e^{i \mathbf{q}_2 \cdot \textbf{r}_i} + \mathbf{u}_3 e^{i \mathbf{q}_3 \cdot \textbf{r}_i},
\end{equation}
where the vectors $\mathbf{u}_{\ell}$ must satisfy
\begin{equation}
\begin{split}
&\mathbf{u}_1 ^2 + \mathbf{u}_2 ^2 + \mathbf{u}_3 ^2 = 1 \\
&\mathbf{u}_i \cdot \mathbf{u}_j = 0 ~~~(i \neq j),\\
\end{split}
\end{equation}
so that $|\mathbf{S}_i|=1$ for all spins. We note that for the wavevectors $\mathbf{q}_{\ell}$ describing type-I order (and also type-II order, as we will see in the next subsection), $e^{i \mathbf{q}_{\ell} \cdot \textbf{r}_i} = \pm 1$ for all sites $\mathbf{r}_i$ of the fcc lattice.
Plugging the ansatz Eq.~\eqref{eq:spins_I} into the Hamiltonian Eq.~\eqref{eq:H1} to compute the expectation value of the anisotropic terms, we find that only the Kitaev term contributes to the classical energy per site:
\begin{equation}
    \label{eq:2}
\Delta E_{\text{I}} = K \left(-1 + 2\left((u_1^{x})^{2} + (u_2^{y})^{2} + (u_3^{z})^{2}\right)\right).
\end{equation}

By minimizing $\Delta E_{\text{I}}$ with respect to $\mathbf{u}_{1}$, $\mathbf{u}_{2}$ and $\mathbf{u}_{3}$ with the constraint $|\mathbf{S}_i|=1$, it is clear from Eq.~\eqref{eq:2} that one can find single-q, a double-q and a triple-q solution which all minimize $\min \Delta E_{\text{I}} = -|K|$. Therefore, the accidental degeneracy of the ground state is not fully lifted: the anisotropic model has a continuous degenerate ground state manifold of type-I states, for which the $\mathbf{u}_{\ell}$ are shown in Table~\ref{tab:I_Kitaev}. 

\begin{table}[htbp]
\centering
\begin{tabular}{|c|c|c|}
\hline
Sign of $K$ & $K<0$    & $K>0$ \\
\hline
\hline
$\min \Delta E_{\text{I}}$ & $+ K$ & $- K$ \\
\hline
\hline
single-q & $\begin{aligned}[c] \mathbf{u}_1 &= \xi_1 \left(1,0,0 \right) \\ \mathbf{u}_2 &=\mathbf{u}_3 = \mathbf{0} \end{aligned}$
&$\begin{aligned}[c] \mathbf{u}_1 &= \left(0,\cos \alpha, \sin \alpha \right) \\ \mathbf{u}_2 &=\mathbf{u}_3 =\mathbf{0} \end{aligned}$ \\
\hline
double-q & $\begin{aligned}[c] 
\mathbf{u}_1 &= \xi_1 \left(u_1,0,0 \right) \\
\mathbf{u}_2 &= \xi_2 \left(0,u_2,0\right) \\
\mathbf{u}_3 &= \mathbf{0} \\
(u_1^2 &+ u_2^2 = 1)
\end{aligned}$
& $\begin{aligned}[c] 
\mathbf{u}_1 &= \xi_1 \left(0,u_1,0 \right) \\
\mathbf{u}_2 &= u_2 \left(\cos \alpha,0,\sin \alpha \right) \\
\mathbf{u}_3 &= \mathbf{0} \\
(u_1^2 &+ u_2^2 = 1)
\end{aligned}$ \\
\hline
triple-q & $\begin{aligned}[c] 
\mathbf{u}_1 &= \xi_1 \left(u_1,0,0 \right) \\
\mathbf{u}_2 &= \xi_2 \left(0,u_2,0 \right) \\
\mathbf{u}_3 &= \xi_3 \left(0,0,u_3 \right) \\
(u_1^2 &+ u_2^2+ u_3^2 = 1) \\
\end{aligned}$
& $\begin{aligned}[c] 
\mathbf{u}_1 &= \xi_1 \left(0,u_1,0 \right) \\
\mathbf{u}_2 &= \xi_2 \left(0,0,u_2 \right)\\
\mathbf{u}_3 &= \xi_3 \left(u_3,0,0 \right)\\
(u_1^2 &+ u_2^2+ u_3^2 = 1) \\
\end{aligned}$ \\
\hline
\end{tabular} 
\caption{Configurations of the $\mathbf{u}_{\ell}$ vectors in type-I ground states selected by Kitaev anisotropy, separately for cases $K >0$ and $K <0$. $\alpha$ is an arbitrary angle and $\xi_{\ell} = \pm 1$.
For example, in the single-q ground state with wavevector $\mathbf{q}_1$, the spins alternate antiferromagnetically in the [100] direction with which they are collinear (resp.\ orthogonal) for $K<0$ (resp. $K>0$).}
\label{tab:I_Kitaev}
\end{table}
 
The degeneracy must therefore be broken by other mechanisms. In the absence of lattice distortions, order-by-disorder, either thermal or quantum, is usually thought to stabilize collinear states \cite{henley_ordering_1987}, so that we expect that the single-q states will be selected and multi-q states will be unstable. We return to this in Sec.~\ref{sec:fluctuations}, where we show that small anisotropy indeed selects single-q states, but that the situation is more complex when anisotropy is significant.

\subsection{Degeneracy in the type-II phase}

We now consider a region of parameter space with type-II ground state order. The spins can be parametrized as:
\begin{equation}
\textbf{S}_{i} = \mathbf{u}_0 e^{i \mathbf{q}_0 \cdot \textbf{r}_i} + \mathbf{u}_1 e^{i \mathbf{q}_1 \cdot \textbf{r}_i} + \mathbf{u}_2 e^{i \mathbf{q}_2 \cdot \textbf{r}_i} + \mathbf{u}_3 e^{i \mathbf{q}_3 \cdot \textbf{r}_i},
\end{equation}
with ordering wavevectors $\mathbf{q}_0 = (\pi, \pi, \pi)/2$, $\mathbf{q}_1 = (-\pi, \pi, \pi)/2$, $\mathbf{q}_2 = (\pi, -\pi, \pi)/2$ and $\mathbf{q}_3 = (\pi, \pi, -\pi)/2$. The $\mathbf{u}_{\ell}$ vectors must in this case satisfy:
\begin{equation}
\begin{split}
\mathbf{u}_{0}^{2} + \mathbf{u}_{1}^{2} + \mathbf{u}_{2}^{2} + \mathbf{u}_{3}^{2} &= 1 \\
\mathbf{u}_{0} \cdot \mathbf{u}_{1} + \mathbf{u}_{2} \cdot \mathbf{u}_{3} &= 0 \\
\mathbf{u}_{0} \cdot \mathbf{u}_{2} + \mathbf{u}_{3} \cdot \mathbf{u}_{1} &= 0 \\
\mathbf{u}_{0} \cdot \mathbf{u}_{3} + \mathbf{u}_{1} \cdot \mathbf{u}_{1} &= 0.
\end{split}
\end{equation}
Like in the type-I case, only one of the anisotropic terms, here the Gamma term, contributes to the anisotropy energy:
\begin{equation}
\begin{split}
\Delta E_{\text{II}} = - 2\Gamma ~ (& u_0^{x} u_0^{y} + u_0^{y} u_0^{z} + u_0^{z} u_0^{x} \\
- & u_1^{x} u_1^{y} + u_1^{y} u_1^{z} - u_1^{z} u_1^{x}  \\
- & u_2^{x} u_2^{y} - u_2^{y} u_2^{z} + u_2^{z} u_2^{x}  \\
+ & u_3^{x} u_3^{y} - u_3^{y} u_3^{z} - u_3^{z} u_3^{x} ).
\end{split}
\end{equation}

It is here non-trivial to analytically minimize $\Delta E_{\text{II}}$ with respect to $\mathbf{u}_{\ell}$ with the $|\mathbf{S}_i|=1$ constraint and we therefore resort to a numerical calculation (explicit minimization with constraints). For $\Gamma>0$, we find that the degeneracy is lifted in favor of either a single-q state or a quadruple-q state, both with $\min \Delta E_{\text{II}} = -2\Gamma$, with a minumum reachable energy of $-2\Gamma \times 0.96$ for the (equal-weight) double-q manifold and $-2\Gamma \times 0.89$ for the (equal-weight) triple-q manifold. Through the same argument as above, we again expect fluctuations to select the single-q, collinear, state over the quadruple-q one. In this case, this would for example be the $\mathbf{q}_0$ state, which is composed of ferromagnetic layers of spins stacked antiferromagnetically in the [111] direction, and in which the spins are aligned with the ordering wavevector ($\mathbf{u}_0 = \pm (1,1,1)/\sqrt{3}$). In the $\Gamma <0$ case, there exist single-, double-, triple-, and quadruple-q states which minimize the energy, with $\min \Delta E_{\text{II}} = \Gamma$. Again, this accidental degeneracy is expected to be lifted via order-by-disorder in favor of the single-q state, in which the spins are collinear. More precisely, in this case, the spins are perpendicular to their ordering wavevectors, and there is a remaining U(1) degeneracy associated with a global rotation within the [111] planes.

The type-II single-q state reproduces the spin arrangement which may have been observed in the half-Heusler compound GdPtBi \cite{kreyssig_magnetic_2011, muller_magnetic_2014, muller_magnetic_2015}. Recent work has argued that instead of the Gamma coupling, an easy-plane single-ion anisotropy, forcing the spins to stay in the [111] planes, could also favor this arrangement \cite{sukhanov_magnon_2020}. However, the easy-plane anisotropy was not considered here as it breaks the cubic symmetry, \textit{i.e.}\ it is {\em not} allowed in a non-distorted crystal, and is therefore artificial.

\subsection{Lifted degeneracy in the type-III phase}

We now turn to type-III order, defined by $\mathbf{q}_1= (\pi/2,\pi,0)$, $\mathbf{q}_2= (0,\pi/2,\pi)$, and $\mathbf{q}_3 = (\pi,0,\pi/2)$ and their opposites. Using Eq.~(\ref{eq:general_S_i}) with $\mathbf{u}_{\ell} = \mathbf{v}_{\ell}-i\mathbf{w}_{\ell}$, the spins can be expressed as:
\begin{equation}
\textbf{S}_{i} = \sum_{\ell=1}^{3} \mathbf{v}_{\ell} \cos(\mathbf{q}_{\ell} \cdot \mathbf{r}_i) + \mathbf{w}_{\ell} \sin(\mathbf{q}_{\ell} \cdot \mathbf{r}_i),
\end{equation}
and the conditions $|\mathbf{S}_{i}| = 1$ become the following geometrical constraints:
\begin{equation}
\begin{split}
& \mathbf{v}_{\ell} \cdot \mathbf{v}_{k} = \mathbf{w}_{\ell} \cdot \mathbf{w}_k = \mathbf{v}_{\ell} \cdot \mathbf{w}_k = 0 ~~~~  \text{if    } \ell \neq k, \\
& \mathbf{v}_{\ell}^{2} = \mathbf{w}_{\ell}^{2} ~~~~ \text{and}~~~~ \sum_{\ell=1}^{3} \mathbf{v}_{\ell}^{2} = \sum_{\ell=1}^{3} \mathbf{w}_{\ell}^{2} = 1.
\end{split}
\end{equation}
The anisotropy energy per site reads:
\begin{equation}
\begin{split}
& \Delta E_{\text{III}} = -4 D \left( (\mathbf{v}_1 \times \mathbf{w}_1)^{x} + (\mathbf{v}_2 \times \mathbf{w}_2)^{y} +(\mathbf{v}_3 \times \mathbf{w}_3)^{z} \right) \\
& - K \left( (v_1^{x})^{2} +(v_2^{y})^{2} + (v_3^{z})^{2} +(w_1^{x})^{2} +(w_2^{y})^{2} + (w_3^{z})^{2} \right).
\end{split}
\end{equation}

\begin{table}[b]
\begin{center}
\begin{tabular}{|c|c|c|}
\hline 
\begin{tabular}{c}
Equal-weight \\
manifold
\end{tabular}
  & $\min \Delta E_{\text{III}}$ & Parametrization \\
\hline
\hline
single-q & $-4D$ & $\begin{aligned}[c] 
\mathbf{v}_3 &= (\cos \alpha,\sin \alpha,0)\\
\mathbf{w}_3 &= (-\sin \alpha, \cos \alpha,0) \\
\mathbf{v}_1 &= \mathbf{w}_1 = \mathbf{v}_2 = \mathbf{w}_2 = 0 \\
\end{aligned}$ \\
\hline 
double-q & $-2D$ & $\begin{aligned}[c] 
\mathbf{v}_3 &= (\cos \alpha,\sin \alpha,0)  / \sqrt{2}  \\
\mathbf{w}_3 &= (-\sin \alpha, \cos \alpha,0) / \sqrt{2}  \\
\mathbf{v}_1 &= \xi_1 (0,0,1)  / \sqrt{2} \\
\mathbf{w}_1 &= \chi_1 (0,0,1) / \sqrt{2}  \\
\mathbf{v}_2 &= \mathbf{w}_2 = 0 \\
\end{aligned}$ \\ 
\hline 
triple-q & $0$ & $\begin{aligned}[c] 
\mathbf{v}_1 &= \xi_1 \mathcal{R} \cdot (1,0,0) / \sqrt{3} \\
\mathbf{v}_1 &= \chi_1 \mathcal{R} \cdot (1,0,0) / \sqrt{3} \\
\mathbf{v}_2 &= \xi_2 \mathcal{R} \cdot (0,1,0) / \sqrt{3} \\
\mathbf{v}_2 &= \chi_2 \mathcal{R} \cdot (0,1,0) / \sqrt{3} \\
\mathbf{v}_3 &= \xi_3 \mathcal{R} \cdot (0,0,1) / \sqrt{3} \\
\mathbf{v}_3 &= \chi_3 \mathcal{R} \cdot (0,0,1) / \sqrt{3} \\
\end{aligned}$ \\ 
\hline 
\end{tabular}
\end{center}
\caption{Lifting the degeneracy of the type-III states with the DM interaction ($D>0$, $K=0$). The selected ground state is a single-q state in which the $\mathbf{v}$ and $\mathbf{w}$ vectors are orthogonal. $\alpha$ is an arbitrary angle, $\mathcal{R}$ is an arbitrary $3\times3$ orthogonal matrix and $\xi_{\ell}, \chi_{\ell} = \pm 1$.}
\label{tab:III_DM}
\end{table}

We first focus on the case $K=0$ and study the effect of the DM coupling. We find that the accidental degeneracy is lifted in favor of a single-q, but {\em non-collinear} ground state, with $ \min \Delta E_{\text{III}}(K=0) = -4 D$ (Table~\ref{tab:III_DM}). The expression of the spins in a single-q state with wavevector $\mathbf{q}_3 = (\pi,0,\pi/2)$ is: 
\begin{equation}
\mathbf{S}_i = \mathbf{v}_3 \cos(\mathbf{q}_3 \cdot \mathbf{r}_i) + \mathbf{w}_3 \sin(\mathbf{q}_3 \cdot \mathbf{r}_i),
\end{equation}
with the constraint $|\mathbf{v}_3| = |\mathbf{w}_3| = 1$. The DM energy in this configuration is $\Delta E_{\text{III}} = -4 D (\mathbf{v}_3 \times \mathbf{w}_3)_{z}$, which reaches its minimum $-4D$ when $\mathbf{v}_3$ and $\mathbf{w}_3$ are orthogonal and lie in the $xy$ plane, for example $\mathbf{v}_3 = \hat{\mathbf{e}}_{x}$, $\mathbf{w}_3 = \hat{\mathbf{e}}_y$.
This state is made of antiferromagnetic [001] layers of spins which lie in the $xy$ plane (all spins in each layer are collinear), which are stacked orthogonally in the [001] direction. Besides the three-fold degeneracy due to cubic symmetry, this ground state has a U(1) degeneracy because a global rotation around the $z$-axis leaves the cross product $(\mathbf{v}_3 \times \mathbf{w}_3)_{z}$ invariant. Note that, even in this minimum-energy configuration, the DM interaction is not fully minimized (it is ``frustrated'').

\begin{figure*}[htbp]
\centering
\includegraphics[width=\textwidth]{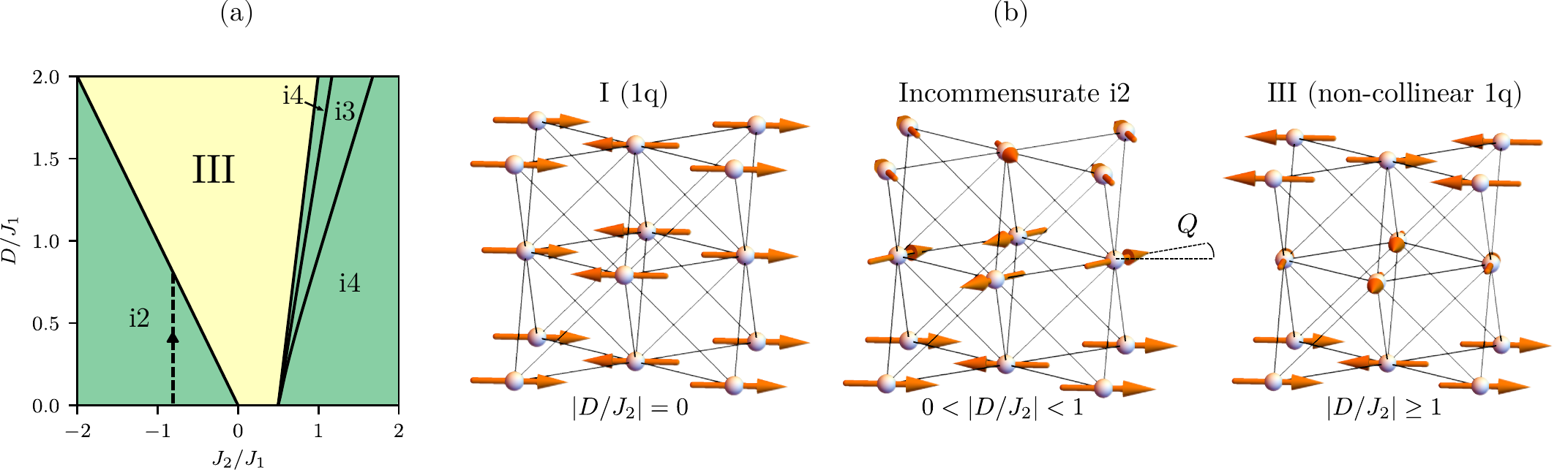}
\caption{Transition from a single-q type-I state to a non-collinear single-q type-III state upon increasing the strength of the DM interactions $D$ (with fixed $J_2 <0$). (a) Path (dotted line) followed in the phase diagram upon showing the states in (b). In the intermediate state i2, the spiral wavevector is $\mathbf{q}=(\pi,0,Q)$ with $Q=\arcsin(D/J_2)$. (b) The I-III phase transition can be seen as a continuous rotation of the spins in the ground state, $z$ layer by $z$ layer. $Q$ is the angle between neighboring layers.}
\label{fig:transition_I_III}
\end{figure*}

From our Luttinger-Tisza analysis, we learned that starting from $J_2 < 0$ and increasing $D$ from zero to a finite value, (dotted line in the diagram Fig.~\ref{fig:transition_I_III}(a)), the system transitions between phase I and phase III, via an incommensurate spiral phase. This incommensurate spiral phase is parametrized by ordering wavevectors of the form $\mathbf{q} = (\pi,0,Q)$ where $Q=\arcsin(D/J_2)$. In particular, the type-III state is reached when $D=-J_2$ such that $Q=\pi/2$. We can now formulate a scenario for this evolution: starting from a single-q type-I state, the DM interactions allows one to lower the energy by a continuous rotation of the spins, layer by layer, until the arrangement reaches the single-q non-collinear type-III state (Fig.~\ref{fig:transition_I_III}(b)).

One can also show that the Kitaev interaction alone ($D=0$, $K\neq 0$) does not lift the degeneracy between single-q and multi-q states, similar to the case of the type-I AFM. In Table~\ref{tab:III_K} we parametrize the single-q, the double-q and the triple-q ground state manifolds of type-III order with finite $K$.
\begin{table}[htbp]
\centering
\begin{tabular}{|c|c|c|}
\hline
Sign of $K$ & $K<0$    & $K>0$ \\
\hline
\hline
$\min \Delta E_{\text{III}}$ & $0$ & $-2 K$ \\
\hline
\hline
single-q & $\begin{aligned}[c] 
	\mathbf{v}_1 &= \xi_1 (0, \cos \alpha,\sin \alpha) \\
	\mathbf{w}_1 &= \chi_1 (0, \cos \beta,\sin \beta) \\
	\mathbf{v}_2 &= \mathbf{w}_2 = \mathbf{0} \\
	\mathbf{v}_2 &= \mathbf{w}_3 = \mathbf{0} \\
\end{aligned}$
& $\begin{aligned}[c] 
	\mathbf{v}_1 &= \xi_1 (1,0,0)  \\
	\mathbf{w}_1 &= \chi_1 (1,0,0)  \\
	\mathbf{v}_2 &= \mathbf{w}_2 = \mathbf{0}  \\
	\mathbf{v}_3 &= \mathbf{w}_3 = \mathbf{0} \\
\end{aligned}$ \\
\hline
double-q & $\begin{aligned}[c] 
	\mathbf{v}_1 &= \xi_1 u_1 (0, \cos \alpha,\sin \alpha) \\
	\mathbf{w}_1 &= \chi_1 u_1 (0, \cos \beta,\sin \beta) \\
	\mathbf{v}_2 &= \xi_2 (u_2,0,0) \\
	\mathbf{w}_2 &= \chi_2 (u_2,0,0) \\
	\mathbf{v}_3 &= \mathbf{w}_3 = \mathbf{0} \\
	(u_1^2 &+ u_2^2 = 1)
\end{aligned}$
& $\begin{aligned}[c] 
	\mathbf{v}_1 &= \xi_1 (u_1,0,0) \\
	\mathbf{w}_1 &= \chi_1 (u_1,0,0) \\
	\mathbf{v}_2 &= \xi_2 (0,u_2,0) \\
	\mathbf{w}_2 &= \xi_2 (0,u_2,0) \\
	\mathbf{v}_3 &= \mathbf{w}_3 = \mathbf{0} \\
	(u_1^2 &+ u_2^2 = 1)
\end{aligned}$ \\
\hline
triple-q & $\begin{aligned}[c] 
	\mathbf{v}_1 &= \xi_1 (0, u_1, 0) \\
	\mathbf{w}_1 &= \chi_1 (0, u_1, 0) \\
	\mathbf{v}_2 &= \xi_2 (0,0,u_2) \\
	\mathbf{w}_2 &= \chi_2 (0,0,u_2) \\
	\mathbf{v}_3 &= \xi_3 (u_3,0,0) \\
	\mathbf{w}_3 &= \chi_3 (u_3,0,0) \\
	(u_1^2 &+ u_2^2 + u_3^2= 1)
\end{aligned}$
& $\begin{aligned}[c] 
	\mathbf{v}_1 &= \xi_1 (u_1,0,0)  \\
	\mathbf{w}_1 &= \chi_1 (u_1,0,0)  \\
	\mathbf{v}_2 &= \xi_2 (0,u_2,0)  \\
	\mathbf{w}_2 &= \chi_2 (0,u_2,0)  \\
	\mathbf{v}_3 &= \xi_3 (0,0,u_3) \\
	\mathbf{w}_3 &= \chi_3 (0,0,u_3)  \\
	(u_1^2 &+ u_2^2 + u_3^2= 1)
\end{aligned}$ \\
\hline
\end{tabular} 
\caption{Configurations of the $\mathbf{v}_{\ell}, \mathbf{w}_{\ell}$ vectors in type-III ground states selected by Kitaev anisotropy ($D=0, K\neq 0$). $\alpha, \beta$ are arbitrary angles, and $\xi_{\ell}, \chi_{\ell} = \pm 1$. Note that applying a cubic symmetry transformation to the states described in the table also give a ground state.}
\label{tab:III_K}
\end{table}

\begin{figure}[htbp]
\begin{center}
\includegraphics[width=\columnwidth]{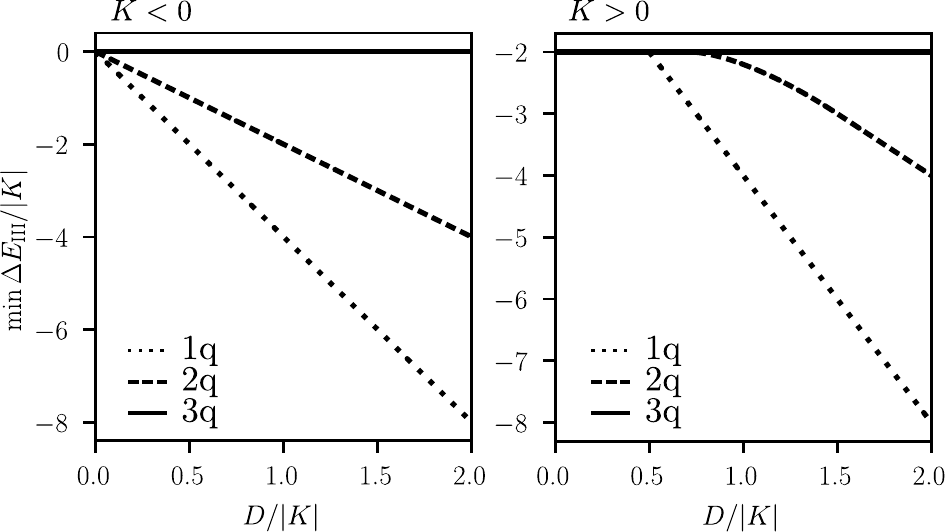}
\caption{Minimal energy of the type-III single-q, double-q and triple-q equal-weight manifolds for $D\neq 0, K\neq 0$.}
\label{fig:III_D_K}
\end{center}
\end{figure}

While the accidental degeneracy of type-III ground states seems to be robust to Kitaev interactions, the DM interaction has a competing effect and tends to lift this degeneracy. In order to investigate the competition between these two effects, we minimized $\Delta E_{\rm III}$ in the presence of both finite $K$ and finite $D$ (Fig.~\ref{fig:III_D_K}). The results show that for $K>0$, the degeneracy is lifted only for $D>K/2$. For $K<0$ however, the degeneracy is lifted for any finite $D$.

\section{Field-induced ground state and selection of non-collinear states}
\label{sec:field}

Having shown in the previous section that an accidental degeneracy persists in multiple regions of the phase diagram even in presence of anisotropic interactions, we expect that this residual degeneracy will be lifted in presence of an external magnetic field $\mathbf{h}$. In this section, we now address the question of the ground state selection by an additional Zeeman energy term $H_{\mathbf{h}}=-\mathbf{h}\cdot \sum_{i}\mathbf{S}_i$, specifically in phases I and II. In particular, we will show that the magnetic field interplays with anisotropic exchange, allowing for the selection of different ground states depending on the direction of the applied field, including non-collinear and non-coplanar states.

For a small magnetic field $\mathbf{h}$, one expects that the spins will uniformly cant in the direction of the field. The spins in the canted state can be written as 
\begin{equation}
\mathbf{S}_i = \mathbf{m} + \mathbf{s}_i,
\end{equation}
where $\mathbf{m}=\chi \mathbf{h}$ is the magnetization per site ($\chi$ is the magnetic susceptibility) and $\mathbf{s}_i$ is the pure antiferromagnetic modulation ($\sum_i \mathbf{s}_i = 0$), which can be expressed as 
\begin{equation}
\mathbf{s}_i = \frac{1}{2} \sum_{\ell} \bigl[\mathbf{u}_{\ell} e^{i\mathbf{q}_{\ell}\cdot \mathbf{r}_i} + 
\mathbf{u}^{*}_{\ell} e^{-i\mathbf{q}_{\ell}\cdot \mathbf{r}_i}\bigr],
\end{equation}
where the $\mathbf{q}_{\ell}$ are the Luttinger-Tisza ordering wavevectors obtained in the {\em zero}-field Luttinger-Tisza approach, and the associated vectors $\mathbf{u}_{\ell}$ are our variational parameters.

We assume that the $g$-tensor is site-independent (so that the effective magnetic field $\mathbf{h}$ is everywhere the product of the scalar $g$-factor and the physical magnetic field). Moreover, we compare the energies of the states with the same $\mathbf{m}$ rather than $\mathbf{h}$: in other words, $\mathbf{m}$ is chosen to be the control parameter rather than the field.

\subsection{Canted type-I}
\label{sec:canted-type-i}

 We first study the case of canted type-I antiferromagnetism, i.e.\ we consider model parameters deep inside the stability region of the type-I phase. In particular, as shown in the previous section, the type-I single-q and multi-q states are degenerate even in the presence of all anisotropic interactions.
In the presence of a net magnetization $\mathbf{m}$ the unit-length conditions $|\mathbf{S}_i|=1$ impose the following geometrical constraints on the variational vectors $\mathbf{u}_{\ell}$:
\begin{equation}
\label{eq:constraints_I_m}
\begin{split}
\mathbf{m}^{2} + \mathbf{u}_{1}^{2} + \mathbf{u}_{2}^{2} + \mathbf{u}_{3}^{2} &= 1 \\
\mathbf{m} \cdot \mathbf{u}_{1} + \mathbf{u}_{2} \cdot \mathbf{u}_{3} &= 0 \\
\mathbf{m} \cdot \mathbf{u}_{2} + \mathbf{u}_{3} \cdot \mathbf{u}_{1} &= 0 \\
\mathbf{m} \cdot \mathbf{u}_{3} + \mathbf{u}_{1} \cdot \mathbf{u}_{2} &= 0.
\end{split}
\end{equation}
For a given magnetization $\mathbf{m}$ one can compare the energies of the canted single-q, double-q and triple-q Ans\"atze:
\begin{equation}
\label{eq:H_canted}
\begin{split}
H\left[\mathbf{S}_i\right] &= \frac{1}{2} \sum_{i,j} \mathbf{S}_{i} A_{ij} \mathbf{S}_{j} - \mathbf{h} \cdot \sum_{i} \mathbf{S}_i \\
&= \frac{1}{2} \sum_{i,j} \mathbf{s}_{i} A_{ij} \mathbf{s}_{j} + \frac{1}{2} \sum_{i,j} \mathbf{m} A_{ij} \mathbf{m}  - N \mathbf{h} \cdot \mathbf{m} \\
&= \frac{1}{2} \sum_{i,j} \mathbf{s}_{i} A_{ij} \mathbf{s}_{j} + (6 J_1 + 3 J_2 + 2 K - \chi^{-1}) N m^{2}, \\
\end{split}
\end{equation}
where we used $\chi\mathbf{h}=\mathbf{m}$ in the last line, and $N$ is the number of sites. The terms linear in $\mathbf{m}$ vanish because the AFM part of the spins satisfies $\sum_j \mathbf{s}_j = \mathbf{0}$:
\begin{equation}
\sum_{i,j} \mathbf{m} A_{ij} \mathbf{s}_{j} = 2(6 J_1 + 3 J_2 + 2K) ~ \mathbf{m} \cdot \sum_j \mathbf{s}_j = 0.
\end{equation}
Since the $m^2$ term in Eq.~\eqref{eq:H_canted} does not depend on the $\mathbf{u}_{\ell}$, the ground state is determined the choice of the vectors $\mathbf{u}_{\ell}$ which minimizes the quantity $\frac{1}{2} \sum_{i,j} \mathbf{s}_{i} A_{ij} \mathbf{s}_{j}$, i.e., the exchange energy of the the AFM modulation. Moreover, since these states are degenerate in the $J_1$-$J_2$ model, only the anisotropic contribution will distinguish the different Ans\"atze: in other words, this problem amounts to minimizing $\Delta E_{\rm I}$ with respect to the $\mathbf{u}_{\ell}$, given the new constraints imposed by the presence of a magnetization.

We choose to focus on the more symmetric case $m_x = m_y$, i.e.\ we let $\mathbf{m}$ vary in the plane which contains the high-symmetry directions [001], [110] and [111] of the crystal (it is then parametrized by its magnitude $m$ and its angle $\theta$ with respect to the [001] axis) and assume that there is no extra spontaneous symmetry breaking. We first focus on the case $K<0$.

\begin{figure}[tbp]
\begin{center}
\includegraphics[width=\columnwidth]{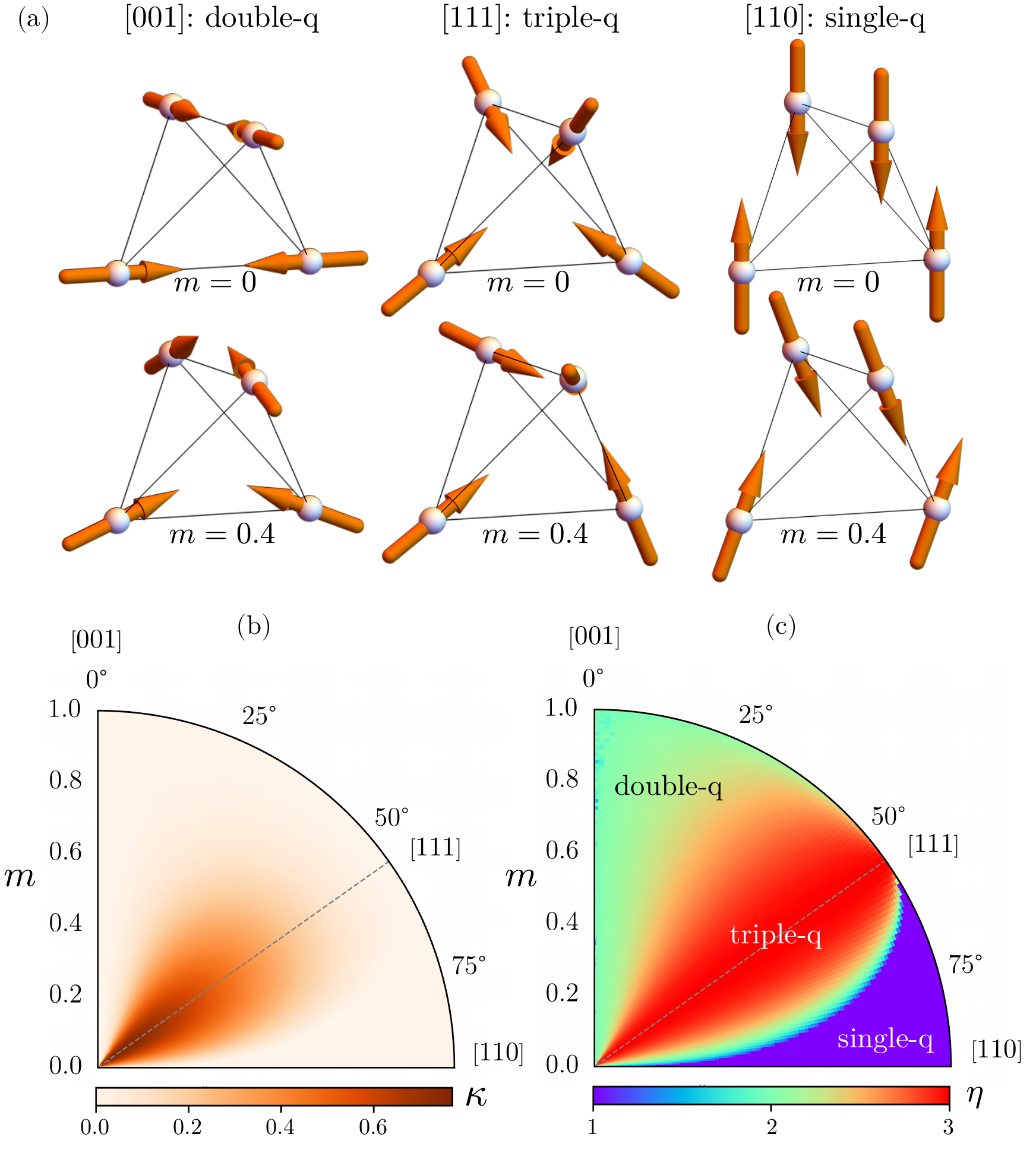}
\caption{Properties of the type-I ground state in the presence of a magnetization and Kitaev coupling $K<0$, as a function of the strength $m$ and the direction of the magnetization $\mathbf{m}$. (a) Spin configurations in the canted double-q, triple-q and single-q states, stabilized when the magnetization is along [001], [111] and [110], respectively. Top row: $m=0$. Bottom row: $m=0.4$. (b) Polar plot of the chirality $\kappa$ (whose square is defined in Eq.~\eqref{eq:chisquared}) per site. (a) Polar plot of the indicator $\eta$ (defined in Eq.~\eqref{eq:eta}).}
\label{fig:field_I}
\end{center}
\end{figure}

Let us first present the selected ground states in the situations when the field is aligned along the different high-symmetry directions: we show them in Fig~\ref{fig:field_I}(a) in cases $m \rightarrow 0$ and $m=0.4$. When the field is along [001] the minimization shows that there is a continuous degeneracy between a canted single-q and a canted double-q ground state:
\begin{equation}
\begin{split}
&\mathbf{u}_1 = w_1 \sqrt{1-m^2} (1,0,0) \\
&\mathbf{u}_2 = w_2 \sqrt{1-m^2} (0,1,0) \\
&\mathbf{u}_3 = \mathbf{0} \\
&(w_1 ^2 + w_2^2 = 1).
\end{split}
\end{equation}
When the field is along [110], the ground state is a canted single-q state:
\begin{equation}
\begin{split}
&\mathbf{u}_{3} = \sqrt{1-m^2} (0,0,1) \\
&\mathbf{u}_1 = \mathbf{u}_2 = \mathbf{0}.
\end{split}
\end{equation}
Finally, when the field is along [111], the ground state is a canted triple-q state with equal weights:
\begin{equation}
\begin{split}
\mathbf{u}_1 &= \sqrt{(1-m^2)/3} (\sqrt{1-2b^2}, -b, -b) \\
\mathbf{u}_2 &= \sqrt{(1-m^2)/3} (-b, \sqrt{1-2b^2}, -b)  \\
\mathbf{u}_3 &= \sqrt{(1-m^2)/3} (-b, -b, \sqrt{1-2b^2}),  \\
\end{split}
\end{equation}
where $b>0$ is such that the constraints Eq.~\eqref{eq:constraints_I_m} are satisfied. In the limit $m\rightarrow 0$ we have $b=0$. This result can be interpreted by using symmetry arguments: for this direction of the field, the system is invariant under $C_3$ symmetry around the [111] axis. The superposition of $\mathbf{q}_1$, $\mathbf{q}_2$ and $\mathbf{q}_3$ modes with equal weights is a good candidate for the ground state because it is invariant under $C_3$, unlike any single-q or double-q states. 

We also carried out the minimization for a generic angle $\theta$ of the magnetization. After finding the optimal parameters $\mathbf{u}_{\ell}$ for each value of $\mathbf{m}$, one can characterize the ground state by a spin chirality $\kappa$, defined by
\begin{equation}
\label{eq:chisquared}
\kappa^2 = \kappa_{[111]}^2 +\kappa_{[\bar{1}11]}^2 +\kappa_{[1\bar{1}1]}^2 +\kappa_{[11\bar{1}]}^2,
\end{equation}
where $\kappa_{\alpha}$ is a chirality around each tetrahedral axis $\alpha$, computed as the sum of the scalar chiralities on triangles lying in the $\alpha$ planes
\begin{equation}
\kappa_{\alpha} = \sum_{\triangle \perp \alpha} \sum_{ijk \in \triangle} \mathbf{S}_i \cdot (\mathbf{S}_j \times \mathbf{S}_k),
\end{equation}
with the three sites $i,j,k$ of the triangle labeled clockwise around the axis $\alpha$. In the case of canted type-I order, the chirality $\kappa$ measures the volume spanned by the three vectors $\mathbf{u}_1, \mathbf{u}_2, \mathbf{u}_3$:
\begin{equation}
\kappa \propto |\mathbf{u}_1 \cdot (\mathbf{u}_2 \times \mathbf{u}_3) |.
\end{equation}
As shown in the polar plot Fig.~\ref{fig:field_I}(b), the chirality $\kappa$ is maximal when $\mathbf{m}$ is directed along $[111]$ and in the limit of vanishing $m$. When $m$ becomes large, the spins are aligned, and $\kappa$ vanishes since it only captures the ``non-coplanarity'' of the arrangement.

While the chirality is a useful quantity to detect a noncoplanar state (triple-q), it fails to detect a coplanar, non-collinear state (double-q) in which the mixed products $\mathbf{S}_i \cdot (\mathbf{S}_j \times \mathbf{S}_k)$ are all zero. We therefore define a quantity $\eta$:
\begin{equation}
\label{eq:eta}
\eta = \frac{(u_1 + u_2 + u_3)^2}{u_1^2 + u_2^2 + u_3^2},
\end{equation}
where $u_{\ell} = |\mathbf{u}_{\ell}|$, which is a continuous quantity measuring the number of ordering wavevectors involved in the AFM ground state, weighted by their amplitude $u_{\ell}$. It is rescaled by the total weight $u_1^2 + u_2^2 + u_3^2 = 1-m^2$. To illustrate the meaning of $\eta$, we note that $\eta = 1$ for a single-q state ($u_1 = 1$, $u_2 = u_3=0$), $\eta=2$ for an equal-weight double-q state ($u_1 = u_2 = \frac{1}{\sqrt{2}}$, $u_3=0$) and $\eta=3$ for an equal-weight triple-q state ($u_1 = u_2 = u_3 = \frac{1}{\sqrt{3}}$). [In a state with non-equal weights, $\eta$ is not an integer.] As shown in the polar plot Fig.~\ref{fig:field_I}(c), $\eta$ increases gradually from the [001] axis ($\eta=2$) to the [111] axis ($\eta=3$), indicating a (smooth) transition from a double-q to a triple-q state. Between [111] and [110] one goes from triple-q ($\eta=3$) to single-q ($\eta = 1$) rapidly.

For a positive Kitaev coupling ($K>0$) the situation is different: the ground state is, for a generic $\mathbf{m}$, a canted single-q state. To show this, we first note that $\Delta E_{\rm I} \geq -K$, and the lower boundary is reached when $u_1^x = u_2^y = u_3^z = 0$. In the presence of finite $\mathbf{m}$, one can always find a single-q state with this property. Indeed, the two conditions $\mathbf{u}_1 \cdot \mathbf{m} =0$ and $u_1^x = 0$ define two planes, whose intersection fixes the direction of $\mathbf{u}_1$. In contrast, it is not always possible to build a double-q state with $u_1^x = u_2^y = 0 = \mathbf{m} \cdot \mathbf{u}_1 = \mathbf{m}\cdot \mathbf{u}_2$, except for $\mathbf{m}$ along the high-symmetry directions [001] or [110], and similarly for the triple-q case. In these two cases we find a degeneracy between single-q, double-q and triple-q ground states. For any other direction the selected state is single-q.

\subsection{Canted type-II}
\label{sec:canted-type-ii}

In the case of canted type-II order, the geometrical constraints $|\mathbf{S}_i| = 1$ lead to:
\begin{equation}
\begin{split}
\mathbf{m} \cdot \mathbf{u}_{\ell} = 0 ~~( \forall\ell = 0,1,2,3) \\
\mathbf{m}^{2} + \sum_{\ell = 0}^{3} \mathbf{u}_{\ell}^{2} = 1 \\
\mathbf{u}_0 \cdot \mathbf{u}_1 +\mathbf{u}_2 \cdot \mathbf{u}_3 = 0 \\
\mathbf{u}_0 \cdot \mathbf{u}_2 +\mathbf{u}_1 \cdot \mathbf{u}_3 = 0 \\
\mathbf{u}_0 \cdot \mathbf{u}_3 +\mathbf{u}_2 \cdot \mathbf{u}_1 = 0. \\
\end{split}
\end{equation}
These equations do not have a triple-q solution in the presence of a finite $\mathbf{m}$, i.e., the triple-q states will not cant (with uniform $\mathbf{m}$) when a magnetic field is added. In contrast, the single-q, double-q and quadruple-q states can lower their energies if the spins cant towards the field. We find that for all values of $\mathbf{m}$, the canted single-q state has a lower energy than the canted double-q and the canted quadruple-q states. In conclusion, in the regions of parameter space where type-II order is stable, the field-induced ground state is a canted single-q state.

\section{Quantum fluctuations and order-by-disorder}
\label{sec:fluctuations}

We now address small quantum fluctuations around the classical ground states. Indeed, as we discussed above, nearest-neighbor anisotropic terms do not fully lift the degeneracies of the AFM states throughout the phase diagram. If no other, larger-scale, interactions exist, then fluctations, thermal or quantum, will lift the degeneracy through the order-by-disorder mechanism \cite{villain_order_1980, henley_ordering_1987, henley_ordering_1989, sheng_ordering_1992, schick_quantum_2020}. At low-enough temperatures, quantum fluctuations will dominate over thermal ones. 

In order to compute the quantum correction to the classical energy, i.e.\ the zero-point energy around the degenerate classical ground states discussed in the previous sections, we proceed within a real space perturbation theory. This approach consists in treating the contribution of the elementary excitations, the magnons, in the quantum Hamiltonian as a perturbative term. This yields a zero-point energy expressed in terms of the classical spins from the classical ground states. In particular, as we show in the following subsection, to second order in perturbation theory the zero-point energy acts as an effective biquadratic interaction between classical spins \cite{jackeli_quantum_2015, zhitomirsky_real-space_2015, larson_effective_2009}. Given the very large magnetic unit cells we address in this work, this real-space perturbative approach is better suited here than the ``conventional'' linear spin wave theory (used for the fcc lattice for example in Refs.~\cite{ader_magnetic_2001} and \cite{yildirim_frustration_1998} for the Heisenberg AFM with type-I and type-II orders respectively, and in Ref.~\cite{li_kitaev_2017} for collinear order in the anisotropic model), which would require the diagonalization of large matrices and numerical integration over the 3d Brillouin zone. 

\subsection{General framework and perturbation theory}

In this subsection, we outline the methodology of the perturbative method used in the calcuation of the zero-point energy. We give more details in Appendix~\ref{sec:perturbation_theory}. We start by defining (real) local orthonormal bases $(\mathbf{\hat{e}}_i^x, \mathbf{\hat{e}}_i^y, \mathbf{\hat{e}}_i^z)$ at each site $i$ of the magnetic unit cell, such that each spin points along its local $z$-axis, $\mathbf{\hat{e}}^z_i$ in the classical ground state. We write the spin operators as:
\begin{equation}
  \label{eq:3} \hat{\mathbf{S}}_i= \hat{\mathsf{S}}_i^z\mathbf{\hat{e}}_i^z+\hat{\mathsf{S}}^x_i\mathbf{\hat{e}}_i^x+\hat{\mathsf{S}}^y_i\mathbf{\hat{e}}_i^y.
\end{equation}
This defines the components of the spin operators in the local bases $\hat{\mathsf{S}}_i^\mu$ (note that we use hats in the notation of the spin operators $\hat{\mathbf{S}}_i$ to distinguish them from the classical moments $\mathbf{S}_i$). The classical ground state is saturated (ferromagnetic) in the local basis, and we have $\langle \hat{\mathbf{S}}_i \rangle = \mathbf{S}_i = S \hat{\mathbf{e}}_i^{z}$, i.e., $\langle \hat{\mathsf{S}}^x_i\rangle=\langle \hat{\mathsf{S}}^y_i\rangle=0$, and $\langle \hat{\mathsf{S}}^z_i\rangle=S$. We write the quantum Hamiltonian as the sum of a mean-field contribution $\hat{\mathcal{H}}_0$ plus that of the quantum contribution of magnons $\delta \hat{\mathcal{H}}$, i.e.\
\begin{equation}
  \label{eq:4}
  \hat{\mathcal{H}} = \frac{1}{2} \sum_{i,j} \hat{\mathbf{S}}_{i} A_{ij} \hat{\mathbf{S}}_{j}  = \hat{\mathcal{H}}_0 + \delta \hat{\mathcal{H}},
\end{equation}
where
\begin{eqnarray}
  \label{eq:5}
  \hat{\mathcal{H}}_0&=& -\sum_i h_i \left(\mathsf{S}^z_i - \frac{1}{2}S\right)\\
  \delta \hat{\mathcal{H}}&=&\frac{1}{2} \sum_{i,j} (\delta \hat{\mathbf{S}}_i^{\perp} + \delta \hat{\mathbf{S}}_i^{\parallel}) A_{ij} (\delta \hat{\mathbf{S}}_{j}^{\perp} + \delta \hat{\mathbf{S}}_j^{\parallel}).
\end{eqnarray}
$\hat{\mathcal{H}}_0$ is expressed in terms of the (classical) local field $\mathbf{h}_i = h_i\mathbf{\hat{e}}_i^z = - \sum_j A_{ij} \mathbf{S}_j$ experienced by the spin at site $i$ in the classical ground state. In the classical states that we consider in this manuscript, the local field has the same magnitude at each site, we denote it $h_0$ in the following. We have split $\delta \hat{\mathcal{H}}$ into longitudinal fluctuations $\delta \hat{\mathbf{S}}^{\parallel} = (\hat{\mathsf{S}}_i^{z} - S)\hat{\mathbf{e}}_i^{z}$ and transverse ones $\delta \hat{\mathbf{S}}^{\perp} = \hat{\mathsf{S}}^x_i\mathbf{\hat{e}}_i^x+\hat{\mathsf{S}}^y_i\mathbf{\hat{e}}_i^y$.

Assuming that the quantum system experiences small quantum mechanical fluctuations around the ground state, we treat $\delta \hat{\mathcal{H}}$ as a perturbation to $\hat{\mathcal{H}}_0$, in the spirit of large-$S$ where $\delta \hat{\mathcal{H}}$ contains an extra multiplicative $1/S$ factor compared to the classical energy.
The first non-zero term in the perturbation theory appears at second order, with expectation value:
\begin{equation}
  \label{eq:6}
  \delta H^{(2)} = \langle 0 | \delta \hat{\mathcal{H}} \mathcal{Q} (E_0 - \hat{\mathcal{H}}_0)^{-1} \delta \hat{\mathcal{H}} | 0 \rangle,
\end{equation}
where $\mathcal{Q}$ is the projector onto the excited states, i.e.\ $\mathcal{Q}=1-|0\rangle\langle0|$, where $|0\rangle$ is the classical ground state $|0\rangle=\otimes_i|S\mathbf{\hat{e}}_i^z\rangle$, and $E_0=-\frac{1}{2}S\sum_ih_i = \frac{1}{2} \sum_{i,j} \mathbf{S}_i A_{ij} \mathbf{S}_j$ is the energy of the classical ground state $|0\rangle$.
In the following, we assume that the local field has the same magnitude $h_0$ on each site, so that we obtain the following zero-point energy:
\begin{equation}
\label{eq:deltaH2}
\delta H^{(2)} = C+\frac{1}{2} \sum_{i,j} \mathbf{S}_{i} \delta \! A_{ij} \mathbf{S}_{j}  -\frac{1}{4 h_0 S^2} \sum_{i,j} (\mathbf{S}_{i} A_{ij} \mathbf{S}_{j})^2
\end{equation}
In Eq.~(\ref{eq:deltaH2}), $C$ is a constant term, i.e.\ it is independent of the ground state,
and $\delta\! A_{ij}$ is a correction to the interaction matrix $A_{ij}$, which renormalizes the coupling constants ($J_{1,2}$, $K$, $\Gamma$, $D$) by a term of order $1/S$. Although these corrections will slightly shift the phase boundaries in the classical phase diagrams, they do not discriminate between different classical ground states away from the boundaries, and we drop them in what follows. Hence, the ground state selection is determined by minimizing the following term:
\begin{equation}
\delta H_{\text{biq}}^{(2)} = -\frac{1}{4 h_0 S^2} \sum_{i,j} (\mathbf{S}_{i} A_{ij} \mathbf{S}_{j})^2.
\end{equation}
$\delta H_{\text{biq}}^{(2)}$ takes the form of an effective biquadratic interaction between pairs of spins, here the square of the classical interaction term with a negative prefactor.
In the following subsections, given a specific region of parameter space for which we know the ground state manifold (minima of $H$), we identify the classical ground state which minimizes $\delta H_{\text{biq}}^{(2)}$.

\subsection{Application to the isotropic model}

Let us first apply the above result to the Heisenberg model with $J_1$ and $J_2$ couplings. In this case, the biquadratic zero-point energy takes the form:
\begin{equation}
\delta H^{(2)}_{\text{biq}} = - \frac{J_1^2}{2 h_0 S^2} \sum_{\langle i,j \rangle} (\mathbf{S}_i \cdot \mathbf{S}_j)^2 - \frac{J_2^2}{2 h_0 S^2}\sum_{\langle\!\langle i,j \rangle\!\rangle} (\mathbf{S}_i \cdot \mathbf{S}_j)^2.
\end{equation}
It becomes clear that this biquadratic interaction term is minimized when the spins are all collinear.

For concreteness, let us consider the region of parameter space $J_1 >0$, $J_2<0$ where type-I order is favored, the zero-point energy per site in a generic type-I state reads
\begin{equation}
\delta H^{(2)}_{\text{biq,I}} = -\frac{J_1^2 S^2}{h_0} \left[ 4 (u_1^{4} + u_2^{4} + u_3^{4}) -1\right] - \frac{3 J_2^2 S^2}{2 h_0} .
\end{equation}
In the equal-weight Ans\"atze we have:
\begin{equation}
\begin{split}
\delta H^{(2)}_{\text{biq,I}}(\text{1q}) &= -\frac{S^2}{h_0}\left(3 J_1^2 + \frac{3}{2} J_2^2\right) \\
\delta H^{(2)}_{\text{biq,I}}(\text{2q}) &= -\frac{S^2}{h_0}\left(J_1^2 + \frac{3}{2} J_2^2 \right) \\
\delta H^{(2)}_{\text{biq,I}}(\text{3q}) &= -\frac{S^2}{h_0}\left(\frac{1}{3} J_1^2 + \frac{3}{2} J_2^2 \right),
\end{split}
\end{equation}
with local field $h_0 = S(4J_1 - 6 J_2)$. [The $J_2$ contribution does not depend on the choice of the ground state because all spins are parallel with their next-nearest neighbors in a type-I state.] Hence, the zero-point fluctuations will select the single-q configurations, since $\delta H^{(2)}_{\text{biq,I}}(\text{1q})$ is smaller than $H^{(2)}_{\text{biq,I}}(\text{2q})$ and $H^{(2)}_{\text{biq,I}}(\text{3q})$. While this effect is likely to persist if small anisotropy is included, for larger anisotropy, however, other ground states may be favored. We explore this in the following subsection.

\subsection{Application to the anisotropic model}

We now take into account the anisotropic couplings in the zero-point energy. According to Eqs.~(\ref{eq:spins_I}) and (\ref{eq:deltaH2}), the biquadratic part of the zero-point energy reads:
\begin{equation}
\label{eq:dHeff_I}
\begin{multlined}
\delta H^{(2)}_{\text{biq}} = -\frac{1}{2 h_0 S^{2}} \sum_{\langle i,j \rangle _{\gamma}} (J_1 \mathbf{S}_{i} \cdot \mathbf{S}_{j} + K S_{i}^{\gamma} S_{j}^{\gamma}  \\ 
+ \Gamma \xi_{ij} (S_{i}^{\alpha} S_{j}^{\beta} + S_{i}^{\beta} S_{j}^{\alpha}) + \mathbf{D}_{ij} \cdot (\mathbf{S}_i \times \mathbf{S}_j) )^{2} \\
- \frac{J_2^{2}}{2h_0 S^2} \sum_{\langle\!\langle i,j \rangle\!\rangle} (\mathbf{S}_i \cdot \mathbf{S}_j)^2.
\end{multlined}
\end{equation}

For simplicity, we now illustrate the consequences of this equation on the type-I states, for which we can explicitly parametrize the spins in the ground state manifold. In the regions of the phase diagram where type-I order is stable, as discussed in Sec.~\ref{sec:degeneracy-type-i}, the single-q and multi-q states are classically degenerate even in the presence of any and all anisotropic couplings. In presence of quantum fluctuations, as we show below, the situation is different. In the following, we consider the case $J_2=0$: as shown in the previous subsection, the $J_2$ contribution to the zero-point energy is uniform in the type-I ground state manifold and thus does not contribute to ground state selection. In this case, the presence of finite anisotropic couplings modifies the norm of the molecular field according to:
\begin{equation}
h_0 = 4S |J_1 - |K||.
\end{equation}

We first focus on the region $K<0$. Using the expression of the $\mathbf{u}_{\ell}$ for the classical ground state manifold obtained in this case (Table~\ref{tab:I_Kitaev}), we find the following zero-point energy per site:
\begin{equation}
\label{eq:dHeff_I_Kneg}
\begin{multlined}
\delta H^{(2)}_{\text{biq,I}} = - \frac{S^2}{h_0} [ ((2 J_1 + K)^2 - 2D^2) (u_1^4 + u_2^4 + u_3^4) \\ + 4 \Gamma^{2}(u_2^2 u_3^2 + u_1^2 u_3^2 + u_1^2 u_2^2) -J_1(J_1+2K)+2D^2].
\end{multlined}
\end{equation}
It is interesting to note that Eq.~(\ref{eq:dHeff_I_Kneg}) shows that the Gamma and DM couplings can lift the degeneracy among type-I states through quantum fluctuations, i.e.\ quantum order-by-disorder, although they do not appear in the energy of type-I states at the classical level. 

We now look for the fluctuation-induced ground state: to this end, we minimize Eq.~(\ref{eq:dHeff_I_Kneg}) with respect to the variational parameters $u_1$, $u_2$ and $u_3$ such that $u_1^2 + u_2^2 + u_3^2 = 1$. 
\begin{figure*}[t]
\centering
\includegraphics[width=\textwidth]{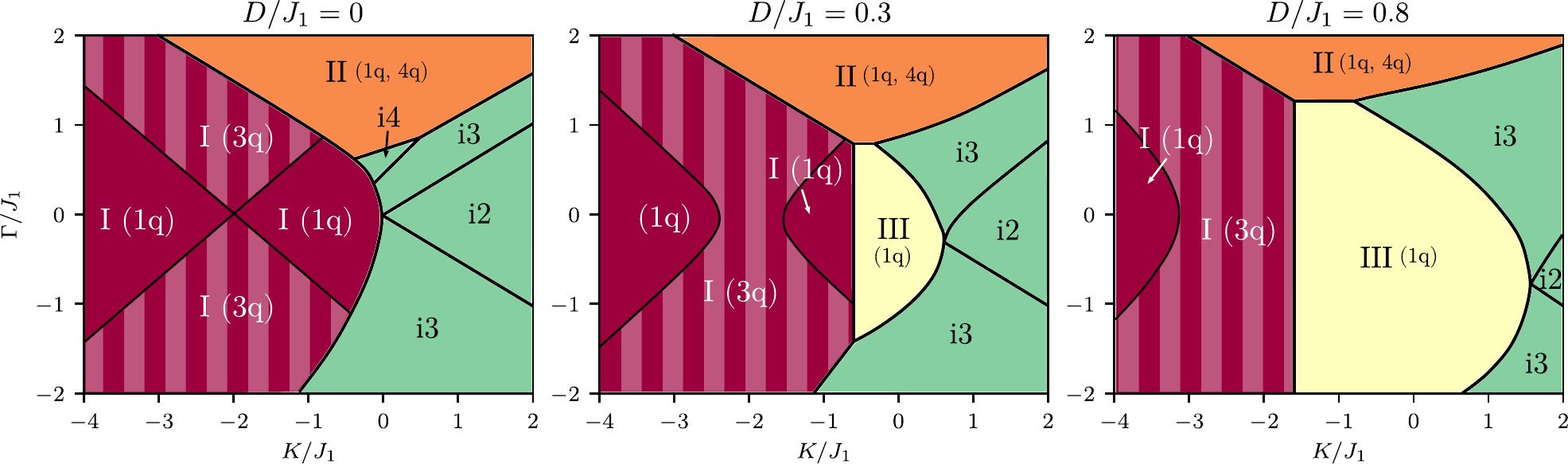}
\caption{ Luttinger-Tisza phase diagram of the model in the $(K,\Gamma)$ plane (with $J_2=0$ and $D$ taking three different values), in which the quantum fluctuations were taken into account {\em for the type-I phase}. The stability regions of the fluctuation-driven triple-q type-I state are indicated by white stripes in the phase diagram.}
\label{fig:diagram_I_quantum}
\end{figure*}
We find that the following equality between coupling constants:
\begin{equation}
\label{eq:degeneracy condition}
(2J_1 + K)^2 - 2D^2 = 2\Gamma^2
\end{equation}
defines a hypersurface in parameter space along which the single-q, the double-q and the triple-q states have the same zero-point energy. In particular, in the case $\Gamma=D=0$ ($J_1$-$K$ model), this corresponds to a degenerate point $K=-2J_1$, and in the case $D=0$ ($J_1$-$K$-$\Gamma$) model this corresponds to a cone-like region in the ($K/J_1,\Gamma/J_1$) plane. In Fig.~\ref{fig:diagram_I_quantum}, we overlay plots of the selected type-I ground state in the ($K,\Gamma$) plane, and of the Luttinger-Tisza phase diagrams which indicate where type-I order is classically stable for three values of $D$.
The hypersurface defined by Eq.~(\ref{eq:degeneracy condition}) separates two regions of parameter space where the degeneracy is lifted by quantum fluctuations.  In the region where $(2J_1 + K)^2 - 2D^2> 2\Gamma^2$ (which includes the small-anisotropy limit $|K|,|\Gamma|, D \ll J_1$), the zero-point energy is minimized by a single-q state (e.g.\ $u_1 = 1$, $u_2 = u_3 =0$), like in the isotropic model. In the region $(2J_1 + K)^2 - 2D^2 < 2\Gamma^2$ however, the zero-point energy is minimized by an equal-weight triple-q state ($u_1 = u_2 = u_3 = 1/\sqrt{3}$). This fluctuation-induced triple-q state is thatwhich we found to arise in the presence of a magnetic field along the [111] axis, in the previous section. For the $K>0$ ground state manifold, the zero-point energy is minimized by a single-q state, for any value of the anisotropic couplings.

We note that in Ref.~\cite{cook_spin-orbit_2015} the authors find, through finite-temperature Monte-Carlo simulations, that {\em thermal} fluctuations select collinear states in the type-I phase even in the presence of $K$, $\Gamma$ anisotropies. This is at odds with our results for quantum fluctuations, and points to an unusual example where, at nonzero temperature, quantum and thermal fluctuations will compete in the ground state selection. 
This effect was recently studied in Ref.~\cite{schick_quantum_2020} in the case of the  {\em nearest-neighbor Heisenberg} model on the fcc lattice. [Recall that this model hosts a degenerate line of spiral ground states -- phase i2 in our notations.] The authors show that fluctuations will lift the line degeneracy in favor of commensurate orders in different ways: quantum flucturations select a (collinear single-q) type-III state, while thermal fluctuations will instead select a (single-q) type-I state.

\section{Conclusion}
\label{sec:conclusion}

In this work we studied the classical spin configurations of an anisotropic nearest- and next-nearest-neighbor fcc antiferromagnet, including symmetry-allowed anisotropic terms, namely the Kitaev, Gamma and Dzyaloshinskii-Moriya interactions. The latter is allowed by the lack of inversion symmetry in the half-Heusler compounds, and was not studied previously in the literature. We found that:
\begin{itemize}
\item The type-I, II and III commensurate antiferromagnetic orders from the isotropic $J_1$-$J_2$ model survive the addition of anisotropy.
\item The accidental single-q/multi-q degeneracy within each of these orders is robust to the Kitaev and Gamma anisotropic coupling terms. In these cases, the anisotropic model ($J_1$-$J_2$-$K$-$\Gamma$ model) hosts a ground-state manifold which includes collinear, non-collinear and non-coplanar configurations.
\item In contrast, the Dzyaloshinskii-Moriya coupling lifts the degeneracy in favor of a non-collinear type-III state.
\end{itemize}
In the regions of parameter space where the degeneracy is not lifted by anisotropic exchange at the classical level (and in particular for type-I order, in which only the Kitaev term contributes classically), we explored further the role of a magnetic field and quantum fluctuations and in particular in which regimes they favored non-collinear magnetic arrangements. 
\begin{itemize}
\item The coupling to a small magnetic field, by explicitly breaking lattice symmetries, will lift the degeneracy. In the case of type-I order, and in the regime $K<0$, collinear (single-q), non-collinear (double-q) and non-coplanar (triple-q) states can be selected, depending on the direction of the field. For $K>0$ a collinear state is selected.
\item Quantum fluctuations will also lift the degeneracy, through the `order-by-disorder' mechanism. For small anisotropy, collinear states are favored. In contrast, in some regimes where anisotropy is significant, a non-coplanar (triple-q) type-I state is selected by minimizing the zero-point energy.
\end{itemize}

The stabilization of multi-q magnetic arrangements may have strong implications on the transport properties of half-Heusler compounds. Most notably, charge carriers can acquire non-trivial Berry phases when coupling to non-coplanar spin textures, leading to an anomalous Hall effect. This Berry phase distribution is enhanced with strong spin-orbit coupling, and may exist even in the case of collinear and coplanar arrangements of spins. From a different viewpoint, non-collinear ordering can lead to Weyl crossings in the electronic bandstructure, which may provide a different origin from that discussed previously to the unusual transport properties (anomalous Hall effect, negative magnetoresistance) observed in some half-Heusler compounds.

\acknowledgements

L.S.\ and S.-S.D.\ acknowledge funding by the European Research Council (ERC) under the European Union's Horizon 2020 research and innovation program (Grant agreement No.\ 853116, acronym TRANSPORT). G.J.\ acknowledges support by the Max-Planck-UBC-UTokyo Centre for Quantum Materials. This research was supported in part by the National Science Foundation under Grant No.\ NSF PHY-1748958.

\bibliographystyle{apsrev4-1}
\bibliography{fcc_magnetism_v17.bib}

\newpage

\begin{appendix}

\section{Ground states of the Heisenberg model}
\label{sec:heisenberg}

\begin{figure}[htbp]
\centering
\includegraphics[width= \columnwidth]{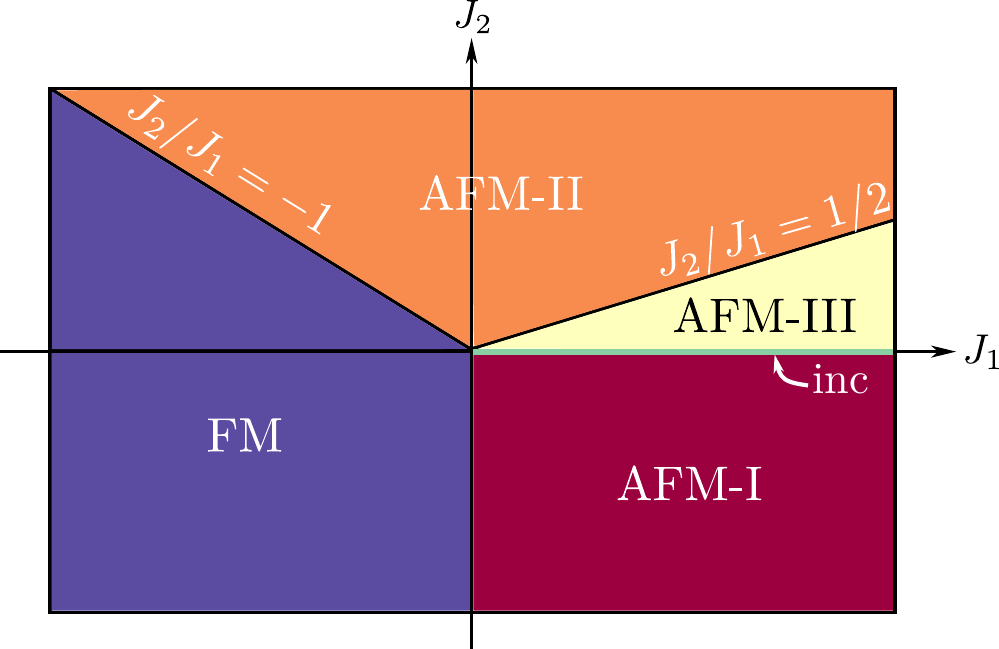}
\caption{Phase diagram of the Heisenberg $J_1$-$J_2$ model on the fcc lattice. It contains a ferromagnetic phase (FM), three commensurate antiferromagnetic (AFM) phases labeled type-I, type-II and type-III orders, as well as an incommensurate (inc) phase with wavevector $(\pi,q,0)$ ($q$ is arbitrary) on the semi-infinite line $J_2=0,J_1>0$. Adapted from Ref.~\onlinecite{yamamoto_spin_1972}.}
\label{fig:J1J2diagram}
\end{figure}

\begin{figure*}[htbp]
\centering
\includegraphics[width=0.9 \textwidth]{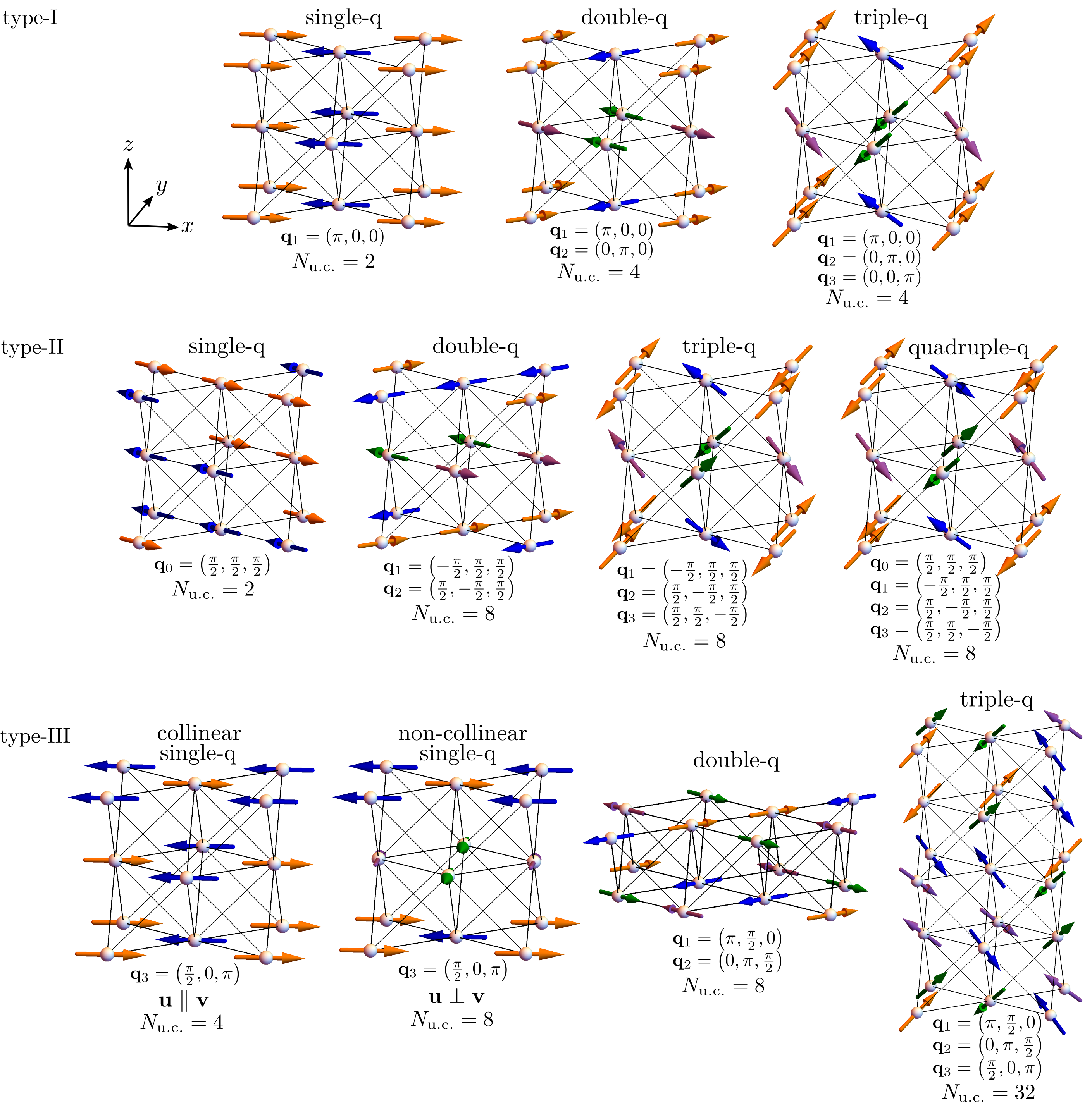}
\caption{Examples of spin arrangements for the fcc antiferromagnetic orders I, II and III stabilized in the $J_1$-$J_2$ phase diagram. The multi-q states shows here are `equal-weight' states. Below each spin configuration, the corresponding ordering wavevectors are written in units of $2/a$, where $a$ is the cubic unit cell parameter, as well as the number of sites in the magnetic unit cell, $N_\text{u.c.}$.}
\label{fig:states_overview}
\end{figure*}

In this section we review the ground states of the Heisenberg model with $J_1$ and $J_2$ couplings, described by the Hamiltonian:
\begin{equation}
H = J_1 \sum	_{\langle i,j \rangle} \mathbf{S}_{i} \cdot \mathbf{S}_{j} + J_2 \sum	_{\langle\!\langle i,j \rangle\!\rangle} \mathbf{S}_{i} \cdot \mathbf{S}_{j}.
\end{equation}
Let us first consider the case $J_1>0$, $J_2 = 0$.
In $\mathbf{q}$ space, all ordering wavevectors of the form $(\pi, q, 0)$, with arbitrary pitch $q$, are ground states. Now, if we additionally consider finite $J_2$ the degeneracy of the AFM states can be split into multiple phases presented in the diagram Fig.~\ref{fig:J1J2diagram}.
Besides the ferromagnetic order, which minimizes the energy for $J_1<0$ and $J_2 < |J_1|$, the diagram features three antiferromagnetic phases labeled type-I, type-II and type-III orders, which differ by their ordering vectors $\mathbf{q}$. In the following three subsections we present these orders in detail and give a parametrization of the spins. In Fig.~\ref{fig:states_overview} we plot examples of single-q and multi-q states, for the three AFM orders I, II and III.

\subsection{Type-I AFM}
Type-I order is defined by the symmetry-related ordering wavevectors $\mathbf{q}_1=(\pi,0,0)$, $\mathbf{q}_2=(0,\pi,0)$ and $\mathbf{q}_3=(0,0,2\pi)$. For a single-q state described by $\mathbf{q}=(\pi,0,0)$, the the spin arrangement is made of ferromagnetic planes stacked antiferromagnetically in the [100] direction. The most general expression of a type-I ground state is:
\begin{equation}
\textbf{S}_{i} = \mathbf{u}_1 e^{i \mathbf{q}_1 \cdot \textbf{r}_i} + \mathbf{u}_2 e^{i \mathbf{q}_2 \cdot \textbf{r}_i} + \mathbf{u}_3 e^{i \mathbf{q}_3 \cdot \textbf{r}_i}.
\end{equation}
In the most general case (i.e. all $\mathbf{u}_{\ell} \neq 0$) the spin configuration is made of 4 ferromagnetic cubic sublattices. By requiring that the length of the spins is $|\mathbf{S}_i|=1$ for the four sublattices, one gets the following geometrical conditions for the $\mathbf{u}_{\ell}$ vectors:
\begin{equation}
\begin{split}
\mathbf{u}_1 ^2 + \mathbf{u}_2 ^2 + \mathbf{u}_3 ^2 = 1 \\
\mathbf{u}_1 \cdot \mathbf{u}_2 = 0 \\
\mathbf{u}_2 \cdot \mathbf{u}_3 = 0 \\
\mathbf{u}_3 \cdot \mathbf{u}_1 = 0. \\
\end{split}
\end{equation}
Single-q type-I order has been in observed in rare-earth half-Heusler compounds CePtBi \cite{wosnitza_magnetic-field-_2006} and NdPtBi \cite{muller_magnetic_2015}.

\subsection{Type-II AFM}
Type-II order is defined by the following symmetry-related ordering wavevectors: $\mathbf{q}_0 = (\pi, \pi, \pi)/2$, $\mathbf{q}_1 = (-\pi, \pi, \pi)/2$, $\mathbf{q}_2 = (\pi, -\pi, \pi)/2$ and $\mathbf{q}_3 = (\pi, \pi, -\pi)/2$. For a single-q state with $\mathbf{q}_0=(\pi,\pi,\pi)/2$, the arrangement is made of ferromagnetic planes of spins stacked antiferromagnetically along the [111] direction. The most general expression of the spins is:
\begin{equation} \label{eq:spins_II}
\textbf{S}_{i} = \mathbf{u}_0 e^{i \mathbf{q}_0 \cdot \textbf{r}_i} + \mathbf{u}_1 e^{i \mathbf{q}_1 \cdot \textbf{r}_i} + \mathbf{u}_2 e^{i \mathbf{q}_2 \cdot \textbf{r}_i} + \mathbf{u}_3 e^{i \mathbf{q}_3 \cdot \textbf{r}_i}
\end{equation}
In this general case the structure is made of 4 cubic N\'{e}el antiferromagnetic sublattices. The conditions $|\mathbf{S}_i| = 1$ give the following equations:
\begin{equation}
\begin{split}
\mathbf{u}_{0}^{2} + \mathbf{u}_{1}^{2} + \mathbf{u}_{2}^{2} + \mathbf{u}_{3}^{2} = 1 \\
\mathbf{u}_{0} \cdot \mathbf{u}_{1} + \mathbf{u}_{2} \cdot \mathbf{u}_{3} = 0 \\
\mathbf{u}_{0} \cdot \mathbf{u}_{2} + \mathbf{u}_{3} \cdot \mathbf{u}_{1} = 0 \\
\mathbf{u}_{0} \cdot \mathbf{u}_{3} + \mathbf{u}_{1} \cdot \mathbf{u}_{1} = 0.
\end{split}
\end{equation}
Single-q type-II order has been measured in compounds GdPtBi \cite{suzuki_large_2016} and TbPtBi \cite{singha_magnetotransport_2019}.

\subsection{Type-III AFM}
Type-III order is defined by $\mathbf{q}_1= (\pi/2,\pi,0)$, $\mathbf{q}_2= (0,\pi/2,\pi)$, and $\mathbf{q}_3 = (\pi,0,\pi/2)$ and their opposites. For a single-q state with wavevector $\pm (\pi,0,\pi/2)$, each spin in any given [100] plane is antiparallel to its nearest neighbor, and the spins on next-nearest [001] planes are antiparallel to one another. 
\begin{equation} \label{eq:spins_III}
\textbf{S}_{i} = \mathbf{u}_1 e^{i \mathbf{q}_1 \cdot \textbf{r}_i} + \mathbf{u}_2 e^{i \mathbf{q}_2 \cdot \textbf{r}_i} + \mathbf{u}_3 e^{i \mathbf{q}_3 \cdot \textbf{r}_i} + \text{c.c.}
\end{equation}
In contrast with phases I and II, here the $\mathbf{u}_{\ell}$ vectors are allowed to have an imaginary part because $\mathbf{q}_{\ell}$ and $-\mathbf{q}_{\ell}$ are not equivalent on the fcc lattice. It is convenient to introduce real vectors $\mathbf{v}_{\ell}$ and $\mathbf{w}_{\ell}$ such that
$\mathbf{u}_{\ell} =  (\mathbf{v}_{\ell} - i \mathbf{w}_{\ell})$. In terms of these vectors, the spins are as parametrised as:
\begin{equation}
\textbf{S}_{i} = \sum_{\ell=1}^{3} \mathbf{v}_{\ell} \cos(\mathbf{q}_{\ell} \cdot \mathbf{r}_i) + \mathbf{w}_{\ell} \sin(\mathbf{q}_{\ell} \cdot \mathbf{r}_i).
\end{equation}
the conditions $|\mathbf{S}_{i}| = 1$ are reduced to the following geometrical constraints:
\begin{equation}
\begin{split}
& \mathbf{v}_{\ell} \cdot \mathbf{v}_{k} = \mathbf{w}_{\ell} \cdot \mathbf{w}_k = \mathbf{v}_{\ell} \cdot \mathbf{w}_k = 0 ~~~~  \text{if    } \ell \neq k, \\
& \mathbf{v}_{\ell}^{2} = \mathbf{w}_{\ell}^{2} ~~~~ \text{and}~~~~ \sum_{\ell=1}^{3} \mathbf{v}_{\ell}^{2} = \sum_{\ell=1}^{3} \mathbf{w}_{\ell}^{2} = 1.
\end{split}
\end{equation}

\section{Details of real space perturbation theory}
\label{sec:perturbation_theory}

In this appendix we consider the quantized version of the Hamiltonian $H$. We show how the contribution of the magnons (excitations around the classical ground state) can be treated as a perturbation in the ground state energy, and we detail the calculation of the energy correction given in equation (\ref{eq:deltaH2}).

For a given ordered ground state, we define a local basis $(\hat{\mathbf{e}}_{i}^{x}, \hat{\mathbf{e}}_{i}^{y}, \hat{\mathbf{e}}_{i}^{z})$ such that $\mathbf{S}_i = S \hat{\mathbf{e}}_{i}^{z}$ is the classical spin at site $i$. The unit vector $\hat{\mathbf{e}}_{i}^{z}$ defines a local quantization axis for the spin. In this local basis, we decompose the spin operator into longitudinal and transverse fluctuations around the classical spin:
\begin{equation}
\hat{\mathbf{S}}_i = \underbrace{ \mathbf{S}_i + \delta \hat{\mathbf{S}}_i^{\parallel}}_{\hat{\mathbf{S}}_i^{\parallel}} + \delta \hat{\mathbf{S}}_i^{\perp} 
\end{equation}
In the above equation, we used hats in the notation of the spin operator to distinguish it from the classical spin vector $\mathbf{S}_i = \langle \hat{\mathbf{S}}_i \rangle$.
More explicitly, we have
\begin{equation}
	\begin{split}
	\hat{\mathbf{S}}_i^{\parallel} &= \hat{\mathsf{S}}_i^{z} \hat{\mathbf{e}}_i^{z} \\
	\delta \hat{\mathbf{S}}_i^{\parallel} &= (\hat{\mathsf{S}}_i^{z} - S) \hat{\mathbf{e}}_i^{z} \\
	\delta \hat{\mathbf{S}}_i^{\perp} &= \frac{1}{\sqrt{2}}( \hat{\mathsf{S}}_i^{-} \hat{\mathbf{e}}_{i}^{+} + \hat{\mathsf{S}}_i^{+} \hat{\mathbf{e}}_{i}^{-})
	\end{split}
\end{equation}
In the above expressions, we have used the following notations: $\hat{\mathsf{S}}^{\pm}_i = \hat{\mathsf{S}}^{x}_i \pm i \hat{\mathsf{S}}^{y}_i$ and $\hat{\mathbf{e}}_{i}^{\pm} = \frac{1}{\sqrt{2}} (\hat{\mathbf{e}}_{i}^{x} \pm i \hat{\mathbf{e}}_{i}^{y})$. The $\hat{\mathsf{S}}^{-}_i$ and $\hat{\mathsf{S}}^{+}_i$ operators lower/raise the projection of the spin at site $i$, creating/destroying a magnon. This process is accompanied by a reduction of the longitudinal component $\hat{\mathsf{S}}^{z}_i$ of the spin. The classical ground state $|0\rangle$ is a saturated state in the local basis characterised by $\hat{\mathsf{S}}^{+}_i |0\rangle = 0$ and $\hat{\mathsf{S}}^{z}_i |0\rangle = S |0\rangle$.
The quantum-mechanical Hamiltonian can be expanded as:
\begin{equation}
\label{eq:quantum_hamiltonian}
\begin{split}
\hat{\mathcal{H}} =& \frac{1}{2} \sum_{i,j} \hat{\mathbf{S}}_{i} A_{ij} \hat{\mathbf{S}}_{j} \\
=&  \underbrace{\frac{1}{2} \sum_{i,j} (\mathbf{S}_i + 2 \delta \hat{\mathbf{S}}_i^{\parallel}) A_{ij} \mathbf{S}_{j}}_{= \hat{\mathcal{H}}_0}\\
 &+ \underbrace{ \frac{1}{2} \sum_{i,j} (\delta \hat{\mathbf{S}}_i^{\perp} + \delta \hat{\mathbf{S}}_i^{\parallel}) A_{ij} (\delta \hat{\mathbf{S}}_{j}^{\perp} + \delta \hat{\mathbf{S}}_j^{\parallel})}_{= \delta \hat{\mathcal{H}}} \\
&+  \underbrace{\sum_{i,j}  \delta \hat{\mathbf{S}}_i^{\perp} A_{ij} \mathbf{S}_j}_{= 0}
\end{split}
\end{equation}
The first term is the mean-field Hamiltonian (easily diagonalized):
\begin{equation}
\hat{\mathcal{H}}_0 = -\sum_{i} h_i  \left( \hat{\mathsf{S}}_{i}^{z} - \frac{1}{2}S \right)
\end{equation}
where we have introduced the norm $h_i = |\mathbf{h}_i|$ of the molecular field, defined as: 
\begin{equation}
\mathbf{h}_i = - \sum_{j \neq i} A_{ij} \mathbf{S}_{j} = - \frac{\partial \hat{\mathcal{H}}}{\partial \hat{\mathbf{S}}_i} \bigg|_{\hat{\mathbf{S}}_j = \mathbf{S}_j}
\end{equation}
Note that in the classical ground state, the spin at site $i$ is aligned with its local field, hence $\mathbf{h}_i = h_i \hat{\mathbf{e}}_{i}^{z}$. 
The energy of the classical ground state reads 
\begin{equation}
E_0 = \langle 0 | \hat{\mathcal{H}} | 0 \rangle = - \frac{1}{2} \sum_{i}h_i S =  \frac{1}{2} \sum_{i,j} \mathbf{S}_{i} A_{ij} \mathbf{S}_{j}.
\end{equation}
which is the classical Hamiltonian $H$ that we have studied so far.
The second term is quadratic in the magnon operators:
\begin{equation}
\delta \hat{\mathcal{H}} = \frac{1}{2} \sum_{i,j} (\delta \hat{\mathbf{S}}_i^{\perp} + \delta \hat{\mathbf{S}}_i^{\parallel}) A_{ij} (\delta \hat{\mathbf{S}}_{j}^{\perp} + \delta \hat{\mathbf{S}}_j^{\parallel})
\end{equation}
Note that the third term vanishes:
\begin{equation}
\sum_{i,j} \mathbf{S}_i \cdot \delta \hat{\mathbf{S}}_j^{\perp}  = -\sum_{j} \mathbf{h}_j \cdot \delta \hat{\mathbf{S}}_j^{\perp} = 0
\end{equation}
because the molecular field is longitudinal (along the local $z$ axis) while the transverse fluctuations are along $x$ and $y$.

Assuming that the quantum-mechanical ground state is close to the classical ground state, we assume that the contribution of magnons $\delta \hat{\mathcal{H}}$ is small, which suggests to treat this term perturbatively. The first order term reads $\langle 0 | \delta \hat{\mathcal{H}} | 0 \rangle = 0$, because the operators $\hat{\mathsf{S}}_j^{+}$ and $(\hat{\mathsf{S}}^{z}_i - S)$ annihilate the saturated (ground) state. The second-order term of the perturbation theory reads:
\begin{equation}
\delta H^{(2)} = \langle 0 | \delta \hat{\mathcal{H}} \mathcal{Q} (E_0 - \hat{\mathcal{H}}_0)^{-1} \delta \hat{\mathcal{H}} | 0 \rangle,
\end{equation}
where $\mathcal{Q} = 1- |0\rangle \langle 0| $ is the projector operator on excited states. The state $\delta \hat{\mathcal{H}} |0 \rangle = \frac{1}{4} \sum_{i,j} (\hat{\mathbf{e}}_{i}^{+} A_{ij} \hat{\mathbf{e}}_{j}^{+}) \hat{\mathsf{S}}_i^{-} \hat{\mathsf{S}}_j^{-} |0 \rangle$ contains pairs of magnons located on two different sites $i$ and $j$ of the lattice, and each of these pairs has a weight given by the matrix element $\hat{\mathbf{e}}_{i}^{+} A_{ij} \hat{\mathbf{e}}_{j}^{+}$ and an excitation energy $h_i + h_j$ above $E_0$. We obtain:
\begin{equation}
\begin{split}
\delta H^{(2)} &= -\frac{1}{16} \sum_{i,j} \frac{|\hat{\mathbf{e}}_{i}^{+} A_{ij} \hat{\mathbf{e}}_{j}^{+}|^{2}}{h_i + h_j} \langle 0 | \hat{\mathsf{S}}_i^{+} \hat{\mathsf{S}}_j^{+} \hat{\mathsf{S}}_i^{-} \hat{\mathsf{S}}_j^{-} |0 \rangle \\
&= -\frac{S^2}{2} \sum_{i,j} \frac{|\hat{\mathbf{e}}_{i}^{+} A_{ij} \hat{\mathbf{e}}_{j}^{+}|^{2}}{h_i + h_j},
\end{split}
\end{equation}
where we used the commutation relation $[\hat{\mathsf{S}}_{i}^{+},\hat{\mathsf{S}}_{i}^{-}] = 2 \hat{\mathsf{S}}^{z}_{i}$. We first see that the quantum fluctuations around the classical state are energetically favorable ($\delta H^{(2)} < 0$). In the following, we assume that the local field $h_i$ has the same norm $h_0$ on each site.

We now need to re-express $\delta H^{(2)}$ in terms of the classical spins $\mathbf{S}_i$. To this end, we recall that $(\hat{\mathbf{e}}^{x}_i, \hat{\mathbf{e}}^{y}_{i},\hat{\mathbf{e}}^{z}_i)$ is an direct orthonormal basis which implies the following equations:
\begin{equation}
\begin{split}
& (\hat{\mathbf{e}}^{x}_i)^{\mu} (\hat{\mathbf{e}}^{x}_i)^{\nu} + (\hat{\mathbf{e}}^{y}_i)^{\mu} (\hat{\mathbf{e}}^{y}_i)^{\nu} + (\hat{\mathbf{e}}^{z}_i)^{\mu} (\hat{\mathbf{e}}^{z}_i)^{\nu} = \delta^{\mu \nu} \\
& (\hat{\mathbf{e}}^{x}_i)^{\mu} (\hat{\mathbf{e}}^{y}_i)^{\nu} - (\hat{\mathbf{e}}^{y}_i)^{\mu} (\hat{\mathbf{e}}^{x}_i)^{\nu} = \epsilon_{\mu \nu \lambda} (\hat{\mathbf{e}}^{z}_i)^{\lambda},
\end{split}
\end{equation}
from which we obtain:
\begin{equation}
2(\hat{\mathbf{e}}^{+}_i)^{\mu} (\hat{\mathbf{e}}^{-}_i)^{\rho} = \delta^{\mu \rho} - (\hat{\mathbf{e}}^{z}_i)^{\mu} (\hat{\mathbf{e}}^{z}_i)^{\rho} -i \epsilon_{\mu \rho \lambda} (\hat{\mathbf{e}}^{z}_i)^{\lambda}.
\end{equation}
This leads to the following rewriting of the $i,j$ term in $\delta E$:
\begin{equation}
\begin{split}
|\hat{\mathbf{e}}_{i}^{+} A_{ij} \hat{\mathbf{e}}_{j}^{+}|^{2} = & \frac{1}{S^4} (\mathbf{S}_{i} A_{ij} \mathbf{S}_{j})^{2} - \frac{1}{S^2} \epsilon_{\mu \rho \lambda} \epsilon_{\nu \sigma \kappa} A_{ij}^{\mu \nu} A_{ij}^{\rho \sigma} S_{i}^{\lambda} S_j^{\kappa}\\
 & - \frac{1}{S^2} (\mathbf{S}_j A_{ji} A_{ij} \mathbf{S}_j) - \frac{1}{S^2} (\mathbf{S}_i A_{ij} A_{ji} \mathbf{S}_i) \\
 &+ \Tr(A_{ji} A_{ij}).
\end{split}
\end{equation}
The first term is the square of the $i,j$ term in the classical energy. The second term contributes a correction $\delta\! A_{ij}$ to the bilinear interaction term $A_{ij}$. The third and fourth terms are single-site quadratic terms of the form $\mathbf{S}_i B_i \mathbf{S}_i$: in Appendix~\ref{sec:single-ion_anisotropy} we show that single-site anisotropies are forbidden by symmetry, and therefore $\mathbf{S}_i B_i \mathbf{S}_i \propto \mathbf{S}_i \cdot \mathbf{S}_i= S^2$ is a global energy shift, independent of the chosen ground state. The fifth term is also a global energy shift.

Hence,
\begin{equation}
\delta H^{(2)} = C+\frac{1}{2} \sum_{i,j} \mathbf{S}_{i} \delta \! A_{ij} \mathbf{S}_{j}  -\frac{1}{4 h_0 S^2} \sum_{i,j} (\mathbf{S}_{i} A_{ij} \mathbf{S}_{j})^2
\end{equation}
where $C$ is a constant, and:
\begin{equation}
\delta\! A_{ij} = \frac{1}{2 h_0} \epsilon^{\mu \rho \lambda} \epsilon^{\nu \sigma \kappa} A_{ij}^{\mu \nu} A_{ij}^{\rho \sigma}.
\end{equation}
We therefore obtain the following corrections to the coupling constants, due to quantum fluctuations:
\begin{equation}
\begin{split}
\delta J_1 &= \frac{1}{h_0} (J_1^2 + J_1 K +  D^2) \\
\delta J_2 &= \frac{1}{h_0}  J_2^2 \\
\delta K &= -\frac{1}{h_0} (J_1 K + \Gamma^2 + D^2) \\
\delta \Gamma &= -\frac{1}{h_0} (D^2 + J_1 K + J_1 \Gamma) \\
\delta D &= -\frac{1}{h_0} D(J_1 - \Gamma). \\
\end{split}
\end{equation}
These corrections are of order $1/S$ with respect to the bare coupling constants: indeed, denoting loosely $J$ the order of magnitude of these coupling constants, we have $h_0 \sim JS$ such that $\delta J/J \sim J/h_0 \sim 1/S$.

\section{Absence of symmetry-allowed single-ion anisotropy}
\label{sec:single-ion_anisotropy}

In this section we study single-ion quadratic terms in the Hamiltonian of the following form:
\begin{equation}
H_a = \sum_{i} \mathbf{S}_i B_{i} \mathbf{S}_i
\end{equation}
where $B_i$ is a $3\times3$ matrix. Similarly to our analysis of the nearest-neighbor quadratic interaction terms, we ask what matrix $B_i$ is allowed by symmetry. To this end, we recall that a point group symmetry transformation can be represented as a $3\times 3$ orthogonal matrix $R$, and that $B_i$ transforms under $R$ as:
\begin{equation}
A_i \rightarrow R^{\rm T} B_i R.
\end{equation}
Symmetry of $H_a$ under the whole point group implies $B_i \propto \mathbf{1}_3$ and thus, no single-ion anisotropy term is allowed.

\end{appendix}

\end{document}